\documentclass[a4paper,fleqn,final]{cas-sc}

\usepackage[section]{placeins}
\usepackage{amssymb}
\usepackage{amsmath}
\usepackage{amsfonts}
\usepackage{upgreek}
\usepackage{siunitx}
\usepackage{listings}
\usepackage{multirow}
\usepackage{multicol}
\usepackage[sort&compress,square,numbers]{natbib}
\usepackage{siunitx}
\usepackage{bigdelim}
\usepackage{color}
\usepackage{stmaryrd}
\usepackage{soul}
\usepackage{booktabs}
\usepackage[section]{placeins} % For '\input{file}'
\usepackage[titletoc]{appendix}
\usepackage{calc}
\usepackage[capitalize]{cleveref}
\usepackage{tabularx}

\usepackage{xspace}

\usepackage{graphicx}
\usepackage{graphics}
\usepackage{float}
\usepackage{caption}
\usepackage{subcaption}
\usepackage{wrapfig}
\usepackage[percent]{overpic}
\usepackage{varwidth}
\usepackage{adjustbox}

\usepackage{tikz}
\usepackage{pgfplots}
\usetikzlibrary{arrows,matrix,positioning,fit}
\usetikzlibrary{shapes,positioning}
\usetikzlibrary{shapes.multipart}
\usetikzlibrary{backgrounds}
\pgfplotsset{compat=1.14}
\usepgfplotslibrary{colorbrewer}
\usepgfplotslibrary{patchplots}
\usepgfplotslibrary[colorbrewer]
\usetikzlibrary{pgfplots.colorbrewer}
\usetikzlibrary[pgfplots.colorbrewer]
\usepgfplotslibrary{fillbetween}
\usepackage{tikz-layers}
\usepgfplotslibrary{units}
\usetikzlibrary{spy}
\usepackage{pgfplotstable}
\graphicspath{ {figs/} }
\newcommand\px[2]{\frac{\partial #1}{\partial {#2}}}

\newcommand{\half}{\frac{1}{2}}

\newlength\myheight
\newlength\mydepth
\settototalheight\myheight{Xygp}
\DeclareRobustCommand\onedot{\futurelet\@let@token\@onedot}
\def\@onedot{\ifx\@let@token.\else.\null\fi\xspace}

\makeatother
\usepackage{acro}

\DeclareAcronym{dsem}{
  long={discontinuous spectral element method},
  short=DSEM
}
\DeclareAcronym{fr}{
  long={flux reconstruction},
  short=FR
}
\DeclareAcronym{kxrcf}{
  long={\citet{Krivodonova2004} sensor},
  short=KXRCF
}
\DeclareAcronym{sbp}{
  long={summation-by-parts},
  short=SBP
}

\begin{document}
\let\WriteBookmarks\relax
\def\floatpagepagefraction{1}
\def\textpagefraction{.001}

\shorttitle{Positivity-preserving discontinuous spectral element methods for compressible multi-species flows}    
\shortauthors{W. Trojak et al.}  
\title[mode = title]{Positivity-preserving discontinuous spectral element methods for compressible multi-species flows}  

% \tnotemark[<tnote number>] 


\author[1]{Will Trojak}[orcid=0000-0002-4407-8956]
\cormark[1]
\ead{w.trojak@ibm.com}
\affiliation[1]{organization={IBM Research Europe},
            addressline={The Hartree Centre}, 
            city={Warrington},
            postcode={WA4 4AD}, 
            state={Cheshire},
            country={UK}}

\author[2]{Tarik Dzanic}[orcid=0000-0003-3791-1134]
\affiliation[2]{organization={Department of Mechanical and Aerospace Engineering},
            addressline={Princeton University}, 
            city={Princeton},
            postcode={08544}, 
            state={NJ},
            country={USA}}

% \author[<aff no>]{<author name>}[<options>]
% \ead{}
% \affiliation[<aff no>]{organization={},
% %          citysep={}, % Uncomment if no comma needed between city and postcode

\cortext[1]{Corresponding author}

\begin{abstract}
We introduce a novel positivity-preserving, parameter-free numerical stabilisation approach for high-order discontinuous spectral element approximations of compressible multi-species flows. The underlying stabilisation method is the adaptive entropy filtering approach (Dzanic and Witherden, \textit{J. Comput. Phys.}, 468, 2022), which is extended to the conservative formulation of the multi-species flow equations. We show that the straightforward enforcement of entropy constraints in the filter yields poor results around species interfaces and propose an adaptive, parameter-free switch for the entropy bounds based on the convergence properties of the pressure field which drastically improves its performance for multi-species flows. The proposed approach is shown in a variety of numerical experiments applied to the multi-species Euler and Navier--Stokes equations computed on unstructured grids, ranging from shock-fluid interaction problems to three-dimensional viscous flow instabilities. We demonstrate that the approach can retain the high-order accuracy of the underlying numerical scheme even at smooth extrema, ensure the positivity of the species density and pressure in the vicinity of shocks and contact discontinuities, and accurately predict small-scale flow features with minimal numerical dissipation.  
\end{abstract}

% \begin{highlights}
% 
% \end{highlights}

\begin{keywords}
Compressible multi-species flows \sep 
High-order \sep 
Discontinuous spectral elements \sep
Positivity-preserving \sep
Entropy filtering \sep
\end{keywords}

\maketitle

\section{Introduction}\label{sec:intro}
    The interaction and mixing of miscible fluids remain extensively studied phenomena, spanning various application areas such as combustion, climate modelling, and chemical process engineering \citep{Dunstan2010,Toor1969,Sharan2021,Kurnia2014}. This behaviour is further complicated in specific applications with high-speed flows as fluid velocities are sufficiently large such that the compressibility effects cannot be ignored. The interaction of these multi-species flows with high Mach number effects, exhibiting features such as shocks, contact discontinuities, rarefaction waves, and species interfaces, poses a significant challenge for numerical modelling approaches. As such, the development of high-fidelity numerical methods for compressible multi-species flows remains an ongoing field of research with the potential for use in many practical applications.

    Various models have been proposed for compressible multi-species flows, with the early work of \citet{Baer1986} presenting some competing approaches and introducing the now widely-adopted method of approximating the evolution of the density of each species along with the total momentum and energy. This conservative formulation can be straightforwardly implemented and is equally valid in both the incompressible, subsonic and compressible, supersonic regimes. While this approach suffers from the deficiency that conservative formulations cannot preserve pressure equilibrium across species interfaces \citep{Abgrall1996,Abgrall2001,Jenny1997,Johnson2020}, its conservative nature ensures that it can accurately predict shock speeds and strong compressibility effects and, as a result, is critical for modelling multi-species flows in the high Mach regime. However, complex flow features in this class of flows present a challenge in developing efficient and accurate numerical approaches for these governing equations.

    For obtaining these high-fidelity approximations of complex fluid flows, high-order discontinuous spectral element methods (DSEM) have grown in popularity, primarily due to their geometric flexibility, arbitrarily high-order accuracy, and compact data structure suited for high-performance computing. These benefits come at the expense of robustness issues, often in the form of numerical instabilities around discontinuous flow features commonly encountered in the approximation of compressible multi-species flows. The use of DSEM for this class of flows has been attempted in the works such as that of \citet{Billet2011}, \citet{Johnson2020}, and \citet{Tonicello2023}, and these approaches broadly rely on the use of additional numerical stabilisation techniques to ensure that the solution remains well-behaved in the vicinity of discontinuities. However, much like in the application of high-order DSEM to single-species gas dynamics and viscous fluid flows, it is notoriously difficult to design numerical stabilisation approaches that guarantee that the solution remains well-behaved while retaining the accuracy of the high-order numerical scheme, particularly so in a computationally efficient manner without the use of problem-dependent tunable parameters. 

    Therefore, this work aims to introduce and validate a robust and efficient numerical approach for simulating both inviscid and viscous compressible multi-species flows in the context of DSEM. This proposed approach is provably robust in that it guarantees that physical constraints such as positivity of density and pressure are satisfied while also retaining the high-order accuracy and scale-resolving capabilities of DSEM in smooth regions of the flow. Furthermore, it can be efficiently implemented on general unstructured meshes and does not require problem-dependent tunable parameters. The underlying stabilisation method of this approach is the entropy filter of \citet{Dzanic2022}, which enforces constraints such as positivity of density and pressure and a local minimum entropy principle through an adaptive filtering procedure. We show that straightforward implementation of the entropy constraints in the filter is exceedingly deficient for multi-species flows, particularly so around species boundaries, and yields poor results. We then present a novel method for adapting the entropy bounds based on the pressure field by utilising the convergence properties of DSEM through the approach of \citet{Krivodonova2004} which significantly increases the accuracy of the stabilisation approach for multi-species flows. This proposed approach is validated and presented across a variety of multi-species problems on both structured and unstructured grids, ranging from simple transport and shock tube problems to inviscid shock-fluid interactions and three-dimensional viscous flow instabilities. 

    The remainder of this paper is organised as follows. In \cref{sec:prelim}, we present some preliminaries on the governing equations and their properties as well as the underlying high-order DSEM. Then, we introduce in-depth the proposed numerical stabilisation approach in \cref{sec:method}, followed by implementation details in \cref{sec:implementation}. The results of the approach as applied to a variety of flow problems are then shown in \cref{sec:valid}, and conclusions are finally drawn in \cref{sec:conclusions}.

\section{Preliminaries}\label{sec:prelim}

\subsection{Governing equations}\label{sec:gov_eq}
    Assuming thermal equilibrium, the governing equations are given as the evolution of $n$ compressible species along with total momentum and energy. We use a species composition based on mass fraction similar to that of \citet{Abgrall1991}. In conservation form, the \textit{inviscid limit} can be represented as 
    \begin{equation}\label{eq:inviscid_system}
        \px{}{t}
        \begin{bmatrix}
        \pmb{\alpha\rho} \\ 
        \rho \mathbf{V} \\ 
        E\end{bmatrix} 
        + \boldsymbol{\nabla}\cdot
        \begin{bmatrix} 
        \mathbf{V}^T\pmb{\alpha\rho} \\ 
        \rho \mathbf{V}\otimes\mathbf{V} + \mathbf{I}P \\ 
        \mathbf{V}(E + P) 
        \end{bmatrix} = 0,
    \end{equation}
    where
    \begin{equation}
        \pmb{\alpha\rho} = \begin{bmatrix} \alpha_0\rho_0 \\ \vdots \\ \alpha_{n-1}\rho_{n-1}\end{bmatrix} \quad \mathrm{and} \quad \rho = \sum^{n-1}_{i=0}\alpha_i\rho_i
    \end{equation}
    are the vector of species densities and the total density, respectively, $\rho \mathbf{V}$ is the total momentum, and $E$ is the total energy. Furthermore, $\rho \mathbf{V}/ \rho$ is the velocity and $P$ is the pressure. To close this system, we utilise the equation of state 
    \begin{equation}
        P = (\overline{\gamma} - 1) \rho e \quad \mathrm{and} \quad \rho e = E - \half\rho\mathbf{V}\cdot\mathbf{V},
    \end{equation}
    where $\rho e $ is the internal energy and $\overline{\gamma}$ is the specific heat ratio of the total mixture. Assuming thermal equilibrium, this specific heat ratio is calculated through the following mixture relation 
    \begin{equation}
        \overline{\gamma} = \frac{\sum^{n-1}_{i=0}\alpha_i\rho_i c_{p,i}}{\sum^{n-1}_{i=0}\alpha_i\rho_i c_{v,i}},
    \end{equation}
    where $c_{p,i}$ and $c_{v,i}$ are the specific heat capacities at constant pressure and constant volume, respectively, of the given species. This contrasts with the approach of \citet{Ton1996} where the thermal equilibrium assumption is removed.

    The inviscid governing equations can be readily extended to their viscous formulation as 
    \begin{equation}\label{eq:full_system}
        \px{}{t}
        \begin{bmatrix}
        \pmb{\alpha\rho} \\ 
        \rho \mathbf{V} \\ 
        E\end{bmatrix} 
        + \boldsymbol{\nabla}\cdot
        \begin{bmatrix} 
        \mathbf{V}^T\pmb{\alpha\rho} \\ 
        \rho \mathbf{V}\otimes\mathbf{V} + \mathbf{I}P \\ 
        \mathbf{V}(E + P) 
        \end{bmatrix} = 
        \begin{bmatrix} 
        0\\ 
        \overline{\mu} \left (\nabla \mathbf{V} + \nabla \mathbf{V}^T\right) - \overline{\mu} \frac{2}{3}\boldsymbol{\nabla}\cdot \mathbf{V} \\ 
        \overline{\mu} \left (\nabla \mathbf{V} + \nabla \mathbf{V}^T\right)\mathbf{V} - \overline{\mu} \frac{\overline{\gamma}}{\overline{Pr}}\nabla e
        \end{bmatrix},
    \end{equation} 
    where $\overline{\mu}$ and $\overline{Pr}$ are the dynamic viscosity and Prandtl number of the mixture, respectively, and $e = \rho e/\rho$ is the specific internal energy. There are several closure modules for the transport coefficients of the mixture \cite{Arrhenius1887,Grunberg1949,Wilke1950}. In this work, we opt for a simple mass fraction weighting, i.e., 
    \begin{equation}
        \overline{\mu} = \frac{1}{\rho}\sum^{n-1}_{i=0}\alpha_i\rho_i\mu_i \quad \mathrm{and} \quad \overline{Pr} = \frac{1}{\rho}\sum^{n-1}_{i=0} \alpha_i\rho_i Pr_i,
    \end{equation}
    where $Pr$ is the Prandtl number as this model gives good agreement with the method of \citet{Wilke1950}.

\subsection{Entropy principles}
    From \citet{Gouasmi2020}, for a specific choice of entropy functional $\sigma$, the inviscid (i.e., hyperbolic) governing equations satisfy an entropy inequality in the form of 
    \begin{equation}
        \px{\rho \sigma}{t} + \nabla\cdot (\mathbf{V}\rho \sigma) \geq 0, 
    \end{equation}
    where
    \begin{equation}
        \rho \sigma = \sum^{n-1}_{i=0}\alpha_i\rho_i\sigma_i
    \end{equation}
    is the mass fraction weighted entropy. The entropy of each species is defined as 
    \begin{equation}\label{eq:entropy_species}
        \sigma_i = \int_0^T \frac{c_{v,i}(\tau)}{\tau}\mathrm{d}\tau - R_i\log(\rho_i),
    \end{equation}
    where $T$ is the temperature and $R$ is the universal gas constant.
    Assuming that $c_v$ does not vary with temperature, this mixture entropy can be expressed as
    \begin{equation}\label{eq:entropy}
        \sigma = \sum^{n-1}_{i=0} c_{v,i}\alpha_i\rho_i\log{\left((\alpha_i\rho_i)^{1-\gamma_i}T\right)},
    \end{equation}
    where
    \begin{equation}
         T = \frac{\rho e}{\sum^{n-1}_{i=0}c_{v,i}\alpha_i\rho_i}.
    \end{equation}
    \citet{Gouasmi2020} showed that for any arbitrary $\mathbf{x}_0$, this entropy (and by extent, all convex functionals of it) satisfies a local minimum entropy principle in the form of 
    \begin{equation}
        \sigma(\mathbf{x}_0, t + \Delta t) \geq \underset{\mathbf{x} \in D_0}{\min}\ \sigma(\mathbf{x}, t + \Delta t)
    \end{equation}
    for all $\Delta t \geq 0$, where $D_0$ is the domain of influence of $\mathbf{x}_0$ over the range $[t, t + \Delta t]$.
    
\subsection{Discontinuous spectral element methods}
    The governing equations as described by \cref{eq:full_system} can be presented in the form of a general conservation law as
    \begin{equation}
        \px{}{t}\mathbf{u} (\mathbf{x}, t) + \boldsymbol{\nabla}{\cdot} \mathbf{F} (\mathbf{u}) = 0,
    \end{equation}
    where $\mathbf{u}$ is the vector of conserved variables and $\mathbf{F}$ is the flux. We present here a very brief overview of the nodal discontinuous spectral element method (e.g., discontinuous Galerkin \citep{Hesthaven2008}, flux reconstruction \citep{Huynh2007}, etc.) as applied to first-order hyperbolic conservation laws, but for a more in-depth description and the extension to second-order systems, the reader is referred to \citet{Hesthaven2008} and \citet{Huynh2007} (and the referenced works therein). 
    
    In this approach, the domain $\Omega$ is partitioned into $N$ elements $\Omega_k$ such that $\Omega = \bigcup_{N_e}\Omega_k$ and $\Omega_i\cap\Omega_j=\emptyset$ for $i\neq j$. With a slight abuse of notation, the discrete solution $\mathbf{u} (\mathbf{x})$ within each element $\Omega_k$ can be formed through a nodal approximation as  
    \begin{equation}
        \mathbf{u} (\mathbf{x}) = \sum_{i\in \mathcal{S}} \mathbf{u}_i \phi_i(\mathbf{x}),
    \end{equation}
    where  $\mathbf{x}_i\ \forall \ i \in \mathcal{S}$ is a set of solution nodes,  $\phi_i(\mathbf{x})$ are their associated nodal basis functions, and $\mathcal{S}$ is the set of nodal indices for the stencil. The basis functions possess the property that $\phi_i(\mathbf{x}_j) = \delta_{ij}$. We utilise the notation that $\mathbf{u}_i = \mathbf{u}(\mathbf{x}_i)$ and that $\mathbb P_p$
    represents a $p$-th order approximation, taken as the maximal order of $\mathbf{u} (\mathbf{x})$. Furthermore, we assume that on the element interfaces $\partial \Omega_k$, there is a set of interface nodes $\mathbf{x}_i \in \partial \Omega \ \forall \ i \in \mathcal{I}$, where $\mathcal{I}$ is a set of nodal indices for the interface stencil that is a subset of the solution nodes (i.e., $\mathcal{I} \subset \mathcal{S}$). Due to the piecewise continuous nature of the solution approximation, there exist two values of the solution at each of these interface nodes, one from the element of interest, denoted as $\mathbf{u}_i^{-}$, and one from the interface-adjacent element, denoted as $\mathbf{u}_i^{+}$.
    
    A general formulation for the approximation of the flux $\mathbf{f}(\mathbf{x})$ in DSEM can be given as a collocation projection of the flux onto the solution nodes augmented with an interface correction term to account for inter-element interactions. 
    \begin{equation}
        \mathbf{f}(\mathbf{x}) = \sum_{i\in \mathcal{S}} \mathbf{F} \left(\mathbf{u}_i \right)\phi_i(\mathbf{x}) +  \sum_{i\in \mathcal{I}} \overline{\mathbf{F}}\left(\mathbf{u}_i^-, \mathbf{u}_i^+, \mathbf{n}_i \right) \cdot \mathbf{n}\overline{\phi}_i(\mathbf{x}).
    \end{equation}
    Here, $\overline{\mathbf{F}}\left(\mathbf{u}_i^-, \mathbf{u}_i^+, \mathbf{n}_i \right)$ denotes some common interface flux dependent on the interior/exterior values of the solution at the interfaces and their respective normal vector $\mathbf{n}_i$, which is commonly computed using exact or approximate Riemann solvers, e.g., \citet{Godunov1959}, \citet{Rusanov1962}, \citet{Roe1981}. Furthermore, $\overline{\phi}_i(\mathbf{x})$ denotes the correction basis function associated with the given interface node $\mathbf{x}_i$, which can be appropriately chosen to recover methods such as the discontinuous Galerkin approach (see \citet{Huynh2007}). With this approximation of the flux, the semi-discrete form of the governing equations, given as
    \begin{equation}\label{eq:semidiscrete}
        \px{}{t} \mathbf{u}_i = -\boldsymbol{\nabla}\cdot \mathbf{f}(\mathbf{x}_i),
    \end{equation}
    can be readily advanced in time using a suitable temporal integration algorithm.

\section{Methodology}\label{sec:method}
    In this section, we introduce the proposed stabilisation method for high-order DSEM approximations of compressible multi-species flows. The underlying mechanisms of this approach rely on the properties of DSEM and the governing equations presented in \cref{sec:prelim}. We assume here that the numerical discretisation is chosen such as to recover the nodal discontinuous Galerkin method \citep{Hesthaven2008} (either directly or via the flux reconstruction method \citep{Huynh2007}), strong-stability preserving (SSP) explicit time stepping is used, and the common interface fluxes are computed using a positivity-preserving entropy-stable Riemann solver. The method is first introduced with respect to inviscid compressible multi-species flows and extensions to their viscous counterparts are later presented. 
    
    \subsection{Adaptive entropy filter}\label{sec:ent_filter}
    To stabilise the solution in the vicinity of discontinuities in a robust manner, \citet{Dzanic2022} introduced a novel method of enforcing discrete constraints on the solution via an adaptive filtering procedure. We present here a brief overview of the entropy filtering approach in the context of the original work in single-species gas dynamics and then show how this method can be extended to the multi-species system as well as what modifications and improvements are necessary.  The goal of the approach is to ensure that 
    \begin{equation}
        \Gamma \left(\mathbf{u}_i\right) > 0
    \end{equation}
    is satisfied for all $i \in S$, where $\Gamma\left(\mathbf{u}\right)$ is some convex constraint functional (or set of functionals) typically related to convex invariants of the system. To ensure robustness, physical constraints were enforced in the form of positivity of density and pressure (i.e., $\Gamma_1\left(\mathbf{u}\right) = \rho$, $\Gamma_2\left(\mathbf{u}\right) = P$). However, simply enforcing these physical constraints was not sufficient to ensure that the solution remains well-behaved in the vicinity of discontinuities. The key mechanism for stabilisation introduced by \citet{Dzanic2022} was a forward-in-time local minimum entropy constraint of the form
    \begin{equation}
        \Gamma_3\left(\mathbf{u}\right) = \sigma - \sigma_{\min},
    \end{equation}
    where $\sigma_{\min}$ is the discrete minimum entropy within the element and its Voronoi neighbours (see \citet{Dzanic2022}, Section 3.2). 
    
    These constraints were enforced through an adaptive modal filter, where the solution was first decomposed into its modal form as
    \begin{equation}
        \mathbf{u}(\mathbf{x}) = \sum_{i \in S} \hat{\mathbf{u}}_i \psi_i(\mathbf{x}),
    \end{equation}
    for some set of modal basis functions $\psi_i(\mathbf{x})$ computed with respect to the unit measure and their associated modes $\hat{\mathbf{u}}_i$. A filtered solution was then computed using a second-order exponential filtering kernel as 
    \begin{equation}
        \tilde{\mathbf{u}}(\mathbf{x}) = \sum_{i \in S} \hat{\mathbf{u}}_i e^{-\zeta p_i^2}\psi_i(\mathbf{x}),
    \end{equation}
    where $\zeta$ is the filter strength and $p_i$ is the maximal order of the basis function $\psi_i(\mathbf{x})$. The filter strength was computed adaptively through an element-wise scalar optimisation problem where the minimum necessary filter strength is sought such that the filtered solution satisfies the constraints discretely, i.e.,
    \begin{equation}
         \zeta = \underset{\zeta\ \geq\ 0}{\mathrm{arg\ min}} \ \ \mathrm{s.t.} \ \  \left [\Gamma_1\left(\tilde{\mathbf{u}}(\mathbf{x}_i) \right) > 0, \ \Gamma_2\left(\tilde{\mathbf{u}}(\mathbf{x}_i) \right) > 0, \ \Gamma_3\left(\tilde{\mathbf{u}}(\mathbf{x}_i) \right) > 0 \  \ \forall \ i \in S\right ].
    \end{equation}
    This optimisation problem can be readily solved using standard root-bracketing approaches, and due to the equivalency of the element-wise mean of discontinuous Galerkin schemes to first-order Godunov methods, a solution to this optimisation problem is guaranteed to exist given the previously mentioned assumptions on the numerical scheme \citep{Zhang2010, Zhang2011, Zhang2011b, Chen2017, Dzanic2022}. 
    
    The extension of this approach to the multi-species system presents additional options for constraints on the solution. First, the physical constraints on the positivity of the (mixture) density and pressure are retained, i.e.,
    \begin{equation}
        \Gamma_1\left(\mathbf{u}\right) = \rho \quad \mathrm{and} \quad \Gamma_2\left(\mathbf{u}\right) = P.
    \end{equation}
    Then, further constraints are introduced for the positivity of the individual species densities as 
    \begin{equation}
        \Gamma_3\left(\mathbf{u}\right) = \alpha_i\rho_i 
    \end{equation}
    for $0 \leq i \leq n-1$. Note that while from an analytical perspective, $\Gamma_1\left(\mathbf{u}\right)$ is redundant given $\Gamma_3\left(\mathbf{u}\right)$, it is advantageous to separate these constraints numerically due to the inclusion of numerical tolerances. This separation is later discussed in \cref{sec:implementation}. Finally, a straightforward application of the entropy filter of \citet{Dzanic2022} can be done by including the entropy constraint as 
    \begin{equation}
        \Gamma_4\left(\mathbf{u}\right) = \sigma - \sigma_{\min},
    \end{equation}
    where $\sigma$ is defined by \cref{eq:entropy}.
    
    However, as will be shown in the numerical experiments, the naive implementation of the entropy constraints for compressible multi-species flows results in exceedingly deficient approximations, particularly around species interfaces. This deficiency stems from the discontinuities introduced by species interfaces, where even flows with smooth mixture density fields may consist of individually discontinuous species density fields which, by \cref{eq:entropy_species} and \cref{eq:entropy}, can cause numerically ill-behaved entropy fields. As a result, excessive numerical dissipation is introduced in the vicinity of species interfaces which, much like contact discontinuities in single-species flows, do not need nearly as much stabilisation as compressive features such as shocks. 
    
    A novelty of the proposed approach is the introduction of a method to modify these entropy constraints such that they can adequately stabilise the solution in regions where it is necessary while retaining high accuracy around species interfaces and contact discontinuities. To retain the benefits of the entropy filtering approach, this modification must maintain its desirable properties, namely the efficiency, robustness, and lack of tunable parameters. To this end, we propose a switch to the entropy bounds inspired by the work of \citet{Gao2023}, such that the constraint can be expressed as
    \begin{equation}\label{eq:switch}
        \Gamma_4\left(\mathbf{u}\right) = 
        \begin{cases}
            \sigma - \sigma_{\min}, \quad \mathrm{if} \ S \geq 1,\\
            0 \quad \quad \quad \quad \ \  \mathrm{else},
        \end{cases}
    \end{equation}
    where $S$ is some element-wise switching function. The goal of this switching function is to be able to isolate regions where the numerical scheme would exhibit numerical instabilities (i.e., shocks) such that the entropy constraints are necessary to ensure a well-behaved solution while neglecting other discontinuous regions such as contact discontinuities and species interfaces. Most importantly, this switching function should be able to achieve this in a way such as to not require problem-dependent tunable parameters. We note here that regardless of the choice of $S$, \textit{the physical constraints are always enforced}, such that the positivity-preserving properties are independent of the switching function and its purpose is purely to increase the accuracy of the numerical scheme around species interfaces.

\subsection{Interface jump convergence sensor}\label{sec:sensor}
    The proposed approach relies on detecting discontinuities in the pressure field to isolate contact discontinuities and species interfaces from shocks in the flow. While many methods exist for sensing discontinuities in the solution, they typically rely on arbitrary tunable parameters or are difficult to extend to high-order DSEM on unstructured grids in a robust and efficient manner. To this end, we choose to formulate the switching function similarly to the method of \citet{Krivodonova2004}, a parameter-free sensor which utilises the convergence properties of discontinuous Galerkin-type methods to isolate discontinuities in the solution within elements. For DSEM, it is known that the magnitude of the jump in the solution (or some derived quantity thereof) across an element interface is dependent on the regularity of the solution. If we define the jump of some derived quantity of the solution $Q$ across $\partial \Omega_i$ as 
    \begin{equation}
        I_i = \int_{\partial \Omega_i} \left(Q_i^+ - Q_i^-\right) \ \mathrm{d}s,
    \end{equation}
    then it is expected that $I_i$ is of at least $\mathcal O(h^{p+1})$, where $h$ is some characteristic mesh length, if the solution is sufficiently smooth in the vicinity of $\Omega_i$. However, if the solution is discontinuous in the vicinity of $\Omega_i$, then $I_i$ would be of $\mathcal O(h)$. By utilising the convergence properties of the pressure field, the switching function can be defined as
    \begin{equation}\label{eq:sensor}
        S_i = \frac{\left|\int_{\partial \Omega_i} \left(P_j^+ - P_j^-\right) \ \mathrm{d}s \right|}{h^{p+1} \| P_j^-\|_{\infty} | \partial \Omega_i|  }.
    \end{equation}
    With this formulation, the switching function can be evaluated and the entropy bounds can be appropriately modified per \cref{eq:switch} at the time of the filter application. The numerical procedure for evaluating the switching function as well as the formulation for computing the characteristic mesh scale for general grids is presented in \cref{sec:implementation}.
    
\subsection{Operator splitting}
    As the proposed scheme is reliant on the well-posedness of a local minimum entropy principle, care must be taken in its extension to the multi-species Navier--Stokes equations. While the single-species counterpart of the viscous flow equations may satisfy a minimum entropy principle for a particular choice of entropy \citep{Tadmor1986}, it is not evident whether this holds for the multi-species case with the given entropy functional nor whether the numerical scheme would satisfy the minimum entropy principle on the element-wise mean in the presence of the viscous dissipation. To use the proposed scheme for the Navier--Stokes equations, two options can be considered. One may simply enforce entropy constraints on the full hyperbolic-parabolic system even with the possibility of there not existing a solution to the element-wise optimisation problem. Note that this option does not negatively impact the positivity-preserving properties and robustness of the scheme as any stable root-bracketing method would converge to the maximal filter strength in the absence of a solution to the optimisation problem. 
    
    The alternative approach, which is considered in this work, is to split the hyperbolic (inviscid) and parabolic (viscous) operators and selectively apply constraints as necessary. The semi-discrete form in \cref{eq:semidiscrete} is represented in terms of the inviscid and viscous flux as 
    \begin{equation}\label{eq:hyppar_eq}
        \partial_t \mathbf{u} + \boldsymbol{\nabla}{\cdot}\left({\mathbf{F}_I(\mathbf{u})} + {\mathbf{F}_V(\mathbf{u})}\right) = 0.
    \end{equation}
    For some arbitrary $n$ and time step $\Delta t > 0$, an intermediate temporal update can be given as
    \begin{equation}
        \overline{\mathbf{u}}^{n+1} = \mathbf{u}^n - \Delta t \boldsymbol{\nabla}{\cdot}\left({\mathbf{F}_I(\mathbf{u}^n)} \right),
    \end{equation}
    corresponding to an inviscid substep of the governing equations. If necessary, the boundary conditions must be appropriately modified to ensure consistency with the governing equations (e.g., no slip wall boundary conditions must be converted to slip wall boundary conditions). As this substep obeys the local discrete minimum entropy principle, the adaptive filtering operation $H({\mathbf{u}})$ can be applied as usual to stabilise the solution where necessary. Afterwards, the full temporal update can be performed by adding the viscous component of the divergence of the flux as
    \begin{equation}
        \mathbf{u}^{n+1} = H(\overline{\mathbf{u}}^{n+1}) - \Delta t \boldsymbol{\nabla}{\cdot}\left({\mathbf{F}_V(\mathbf{u}^n)} \right).
    \end{equation}
    Note that the viscous flux is evaluated in an explicit manner, not through a Strang splitting-type approach, such that the accuracy of the temporal integration method is not detrimentally affected. While the addition of the viscous component typically further stabilises the solution, it is possible in rare scenarios that positivity of density and pressure may not be ensured for $\mathbf{u}^{n+1}$ even though it is for $H(\overline{\mathbf{u}}^{n+1})$. Therefore, a secondary filtering operation is applied to the $\mathbf{u}^{n+1}$ using only positivity constraints to guarantee that the solution remains positivity-preserving. As the violation of the positivity constraints caused by the additional viscous components is very rare \citep{Dzanic2022}, in the vast majority of situations, the secondary filtering operation is not performed. 
    
    The operator splitting approach is presented in terms of a single step of a forward Euler scheme. The approach can naturally be extended to any explicit SSP temporal integration method in a similar manner. For extensions to higher-order SSP schemes as well as a detailed overview of the operator splitting approach, the reader is referred to \citet{Dzanic2022}, Appendix A and \citet{Dzanic2023a}, Section 3.2.
\section{Implementation}\label{sec:implementation}
The proposed numerical scheme was implemented within PyFR \citep{Witherden2014}, a high-order flux reconstruction solver that can be efficiently deployed on massively-parallel GPU and CPU computing architectures. For a given polynomial approximation order $p$, the solution and flux points were distributed along the corresponding Gauss--Legendre--Lobatto quadrature points for tensor-product elements and $\alpha$-optimised \citep{Hesthaven2008} points for simplex elements. Temporal integration was performed using a third-order, three stage SSP Runge--Kutta scheme with a fixed time step. The common inviscid interface fluxes were computed using the HLLC Riemann solver \citep{Ansari2013,Toro2009} as it was found to yield notably better results than approaches such as that of \citet{Rusanov1962}. For the common viscous interface fluxes, the BR2 approach of \citet{Bassi2000} was used. 

To ensure a non-vacuum state for the Riemann solver, a numerical tolerance $\epsilon$ was added to the mixture density and pressure constraints, i.e.,
\begin{equation}
    \Gamma_1(\mathbf{u}) = \rho - \epsilon, \quad \quad \Gamma_2(\mathbf{u}) = P - \epsilon.
\end{equation}
Furthermore, the same tolerance was applied to the entropy constraints as
\begin{equation}
    \Gamma_4(\mathbf{u}) = \sigma - \sigma_{\min} + \epsilon.
\end{equation}
Unless otherwise stated, this tolerance was set as $\epsilon = 10^{-5}$ in this work. For the constraint on the individual species density, the tolerance was not applied through the constraints themselves but instead through adding $\epsilon$ to the initial conditions for each species density component.

Due to the logarithmic nature of \cref{eq:entropy}, the calculation of the entropy is highly prone to numerical precision issues as well as numerically undefined behaviour in the limit of zero species density. The robustness and accuracy of the proposed approach was significantly improved through the use of the modified entropy functional $\sigma^*$, defined as 
\begin{equation}
    \sigma^* = \exp \left [\sum^{n-1}_{i=0} c_{v,i}\alpha_i\rho_i\log{\left(\max (\epsilon, \alpha_i\rho_i)^{1-\gamma_i}T\right)} \right],
\end{equation}
which was used for computing the entropy constraints in place of \cref{eq:entropy}. As the exponential function is strictly convex, this modified entropy retains the local minimum entropy principle of its original form, and the clipping of the species density ensures that the entropy is numerically well-defined. 

Unless otherwise stated, the numerical experiments are assumed to be computed with the entropy switch enabled. The sensor for the entropy switch, defined by \cref{eq:sensor}, was computed discretely across the nodal solution points within each element $\Omega_i$ as 
\begin{equation}
    S_i = \frac{\sum_{j\in S}w_j (P_j^+ - P_j^-)}{h^{p+1}\left (\max_{j\in S}|P_j^-| \right) \left (\sum_{j\in S}w_j \right)},
\end{equation}
where $w_j$ is the corresponding quadrature weight for the given nodal point $\mathbf{x}_j$ within $\Omega_i$. The formulation of the mesh scale $h$ was chosen in such a way to easily generalise between unstructured and structured meshes of varying dimensionality. We define the mesh scale as the diameter of the $d$-ball of equivalent surface area to the element in question, i.e.,
\begin{equation}
    h = \begin{cases}
        A/\pi, \quad\quad \mathrm{if} \ d = 2, \\
        \sqrt{A/\pi}, \quad \, \mathrm{if} \ d = 3,
    \end{cases}
\end{equation}
where $A = \sum_{j\in S}w_j$ is the surface area of $\partial \Omega_i$. While other approaches exist for computing a mesh scale for an unstructured mesh (e.g., inscribed circle/sphere), it was found that the results were not particularly sensitive to how the mesh scale was computed. 

For the adaptive filtering method, the constraints were first checked each time the filtering operation was called, and if the solution satisfied the constraints, no filtering was applied. In the case that the solution violated the constraints, the filter strength was computed using the Illinois root-bracketing approach \citep{Dowell1971} with a maximum of 20 iterations and an early stopping tolerance of $10^{-8}$. A highly-efficient optimisation procedure was utilised which exploits the structured nature of the filtering matrices (see \citet{Dzanic2023a}, Section 4.1), such that the overall cost of the adaptive filtering method was only a small portion of the total compute time. This approach was implemented and deployed on parallel GPU computing architectures with computations performed on up to 40 NVIDIA V100 GPUs.  Further details of the configurations used throughout this work can be found in the electronic supplementary material.
\section{Results}\label{sec:valid}
\subsection{Multi-species Euler equations}
\subsubsection{Near-vacuum convecting density wave}
    As an initial verification of the ability of the proposed approach to recover the high-order accuracy of the underlying DSEM for smooth problems, the convergence of the scheme was evaluated for a smooth convecting density wave. This problem is evaluated on the periodic domain $\Omega=[-0.5, 0.5]\times[-0.05,0.05]$, and the initial conditions are given as
    \begin{subequations}
        \begin{align}
            \rho & = \alpha_0\rho_0 + \alpha_1\rho_1 = \exp \left(-\sigma x^2 \right ) + 4\epsilon, \\
            \alpha_0 &= 1 - \alpha_1 = \half(\sin{(2\pi x)} + 1), \\
            u &= 1, \\
            v &= 0, \\
            P &= 2\epsilon,
        \end{align}
    \end{subequations}
    where $\sigma = 500$ is the strength of the Gaussian wave. We consider the case of two-species, where the specific heat capacities were set as $c_{p,0}=1.4$, $c_{v,0}=1$, $c_{p,1}=4.21$, and $c_{v,1}=2.52$. To make this problem more challenging and to test the robustness of the positivity-preserving properties of the proposed scheme, a near-vacuum state was used for the initial conditions by setting $\epsilon = 10^{-12}$, such that each species density had a minimum value of $2{\cdot}10^{-12}$. The tolerances and minimum density/pressure conditions for the scheme were also modified accordingly. For these extreme conditions, numerical undershoots can easily cause negative density and pressure states which can cause a solver to diverge if it does not explicitly enforce positivity. 

    To evaluate the behaviour of the scheme at smooth extrema, the $L^\infty$ norm of the density error was calculated after one flow-through of the domain. The error and the average rate of convergence for varying approximation orders and mesh resolution is shown in \cref{tab:vacuum_error_kxrcf}. It can be seen that the high-order accuracy of DSEM was not detrimentally affected by the proposed stabilisation method, such that high-order (i.e., $p+1$) convergence was observed even at smooth extrema. However, it may be argued that the observed high-order accuracy for the given problem could be a by-product of the entropy switch as the constant pressure field would not trigger the entropy constraints (i.e., $S \sim 0$) which are primarily responsible for stabilisation in the vicinity of discontinuities. To verify that the stabilisation technique recovers high-order accuracy even with entropy constraints, the convergence tests were repeated with the entropy switch disabled, such that the entropy constraints were always enforced for every element in the domain. The error and convergence rates with the entropy switch disabled are presented in \cref{tab:vacuum_error}. It can clearly be seen that the resulting errors and, by extension, convergence rates, are nearly identical with and without the entropy switch. These results indicate that the proposed stabilisation approach retains the high-order accuracy of the underlying DSEM in smooth regions of the flow while ensuring that physical constraints such as positivity of density and pressure are satisfied by the discrete solution. 

    \begin{table}[tbhp]
        \centering
        \caption{\label{tab:vacuum_error_kxrcf}Convergence in the $L^\infty$ norm of the density error at $t = 1$ with respect to mesh resolution $N$ for the near vacuum convection problem with varying approximation order and \textit{with} the entropy switch. Rate of convergence shown beneath.}
        \begin{tabular}{l|rrrrr}
            \toprule
            $N$ & $\mathbb{P}_1$ & $\mathbb{P}_2$ & $\mathbb{P}_3$ & $\mathbb{P}_4$ & $\mathbb{P}_5$ 
             \\ \midrule
            8 & - & - & \num{0.532241878742436} & \num{0.51950392563647} & \num{0.5307177653915376} \\
            16 & - & \num{0.3068802854100652} & \num{0.06943162805397907} & \num{0.04539071296889796} & \num{0.21411399941955656} \\
            32 & \num{0.31610681226902115} & \num{0.033588427998176185} & \num{0.010963880434086345} & \num{0.0003824779982893878} & \num{0.0007702987403231439} \\
            64 & \num{0.10364963124327908} & \num{0.0028477933300661284} & \num{0.00019119724902649793} & \num{1.1150800004022798e-05} & \num{8.048905185686905e-07} \\
            128 & \num{0.022114141081272742} & \num{0.0003528846161252597} & \num{1.3254357371428327e-05} & \num{5.409267090916714e-07} & - \\ 
            256 & \num{0.0036765970610788123} & \num{4.348406233856128e-05} & \num{1.0570197418457994e-06} & - & - \\ 
            512 & \num{0.0006260359473178045} & \num{5.468239494765825e-06} & - & - & - \\ \midrule
            \textbf{RoC} & \num{2.277710192382107} & \num{3.1621017591612826} & \num{3.931853872208715} & \num{5.173757438673769} & \num{6.611090789893324} \\
            \bottomrule
        \end{tabular}
    \end{table}
    
    \begin{table}[tbhp]
        \centering
        \caption{\label{tab:vacuum_error}Convergence in the $L^\infty$ norm of the density error at $t = 1$ with respect to mesh resolution $N$ for the near vacuum convection problem with varying approximation order and \textit{without} the entropy switch. Rate of convergence shown beneath.}
        \begin{tabular}{l|rrrrr}
            \toprule
            $N$ & $\mathbb{P}_1$ & $\mathbb{P}_2$ & $\mathbb{P}_3$ & $\mathbb{P}_4$ & $\mathbb{P}_5$ 
             \\ \midrule
            8 & - & - & \num{0.532241878742436} & \num{0.51950392563647} & \num{0.5307177515384055} \\
            16 & - & \num{0.30688028512715326}  & \num{0.06943163236839212} & \num{0.04539071453497484} & \num{0.21411405585277865} \\
            32 & \num{0.31613065734539625} & \num{0.03371390165721144} & \num{0.010963882407469677} & \num{0.0003825141296662782} & \num{0.0007702988557855972} \\
            64 & \num{0.10364575428349976} & \num{0.002847814322958886} & \num{0.0001912513433789398} & \num{1.1149192762904825e-05} & \num{1.1064288417683699e-06} \\
            128 & \num{0.022114164463088226} & \num{0.0003528651435552588} & \num{1.3161536916883598e-05} & \num{4.6233203143675183e-07} & - \\ 
            256 & \num{0.0036766140295882987} & \num{4.34939765038278e-05}  & \num{1.0962475218700973e-06} & - & - \\ 
            512 & \num{0.0006260674529865451} &  \num{5.4610235555863085e-06} & - & - & - \\ \midrule
            \textbf{RoC} & \num{2.2777113742820814} &  \num{3.1628093816810323} & \num{3.9252010832300126} & \num{5.219079002750473} & \num{6.473376933929473} \\
            \bottomrule
        \end{tabular}
    \end{table}

\subsubsection{Isentropic Euler vortex}
    The convecting Euler vortex \citep{Shu1998} is an isentropic solution to the Euler equations with an analytic expression for its time evolution. Consequently, it is frequently used as a validation case for numerical solvers. For the extension to multi-species flows, the initial conditions can be adapted as
    \begin{subequations}
        \begin{align}
            \rho_i &= \left(1 - \frac{(\gamma_i - 1)\beta^2M^2}{8\pi^2}\exp(2f)\right)^{1/(\gamma_i - 1)}, \quad \mathrm{for} \quad i\in\{1,2\},\\
            u &= U_0 + \frac{\beta y}{2\pi R}\exp(f), \\
            v &= V_0 - \frac{\beta x}{2\pi R}\exp(f), \\
            P &= \frac{1}{\gamma M^2}\rho^{\gamma}, \\
            \alpha_1 &= \half\sin{\left(\frac{\pi}{L}(y-y_0)\right)} + \half,\\
            f &= \frac{1 - (x-x_0)^2 - (y-y_0)^2}{2R^2}, 
        \end{align}
    \end{subequations}
    defined on the domain $\Omega=[x_0-L,x_0+L]\times[y_0-L,y_0+L]$. Here, the parameters $\beta = 13.5$ denote the strength of the vortex, $R = 1.5$ the radius, $U_0 = 0$, $V_0 = 1$ the advection velocities, and $M = 0.4$ the free-stream Mach number. The domain was centred around $x_0 = y_0 = 0$ with an extent of $L = 10$. 
   \begin{figure}[tbhp]
        \centering
        \subfloat[$N=20^2$ (with entropy switch)]{\adjustbox{width=0.33\linewidth, valign=b}{\includegraphics{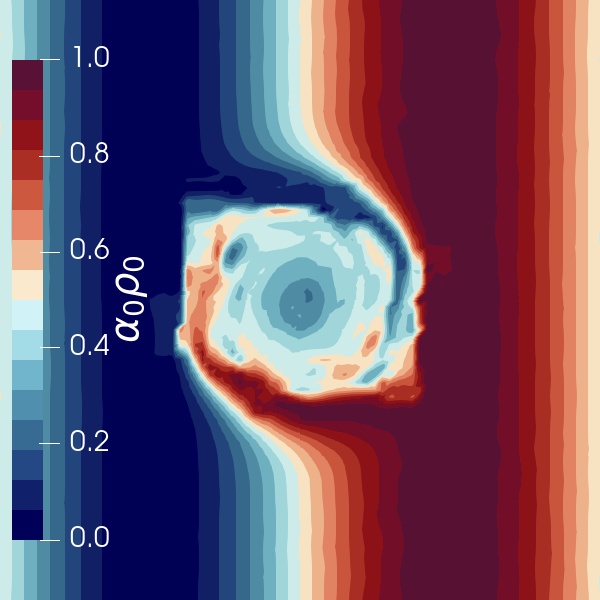}}}
        ~
        \subfloat[$N=40^2$ (with entropy switch)]{\adjustbox{width=0.33\linewidth, valign=b}{\includegraphics{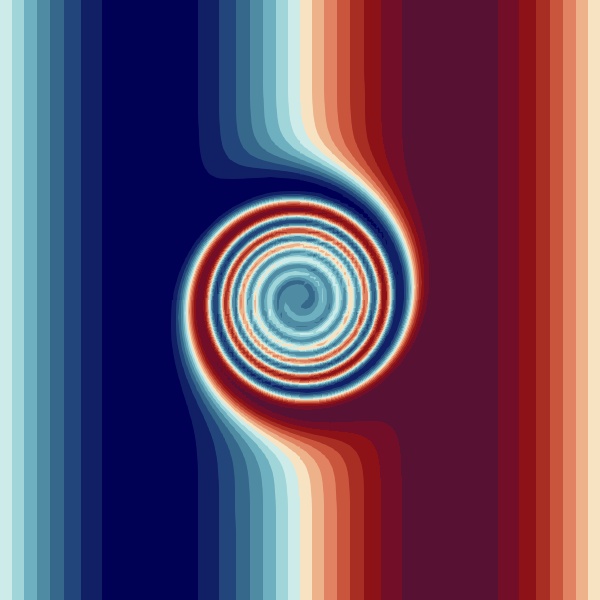}}}
        ~
        \subfloat[$N=80^2$ (with entropy switch)]{\adjustbox{width=0.33\linewidth, valign=b}{\includegraphics{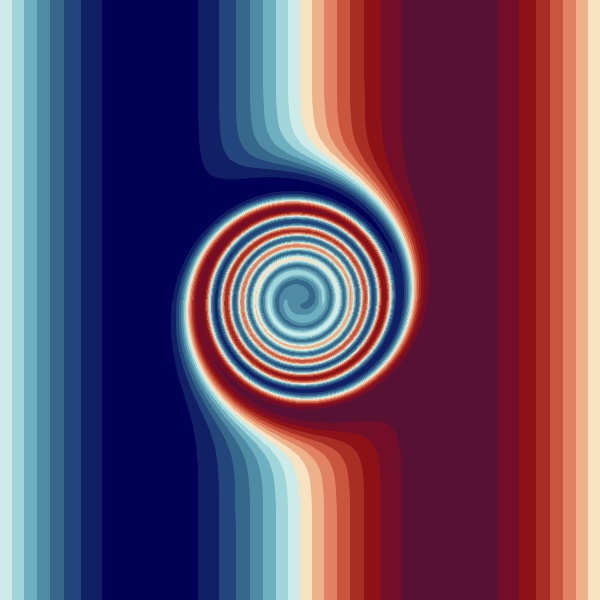}}}\\
        \subfloat[$N=20^2$ (without entropy switch)]{\adjustbox{width=0.33\linewidth, valign=b}{\includegraphics{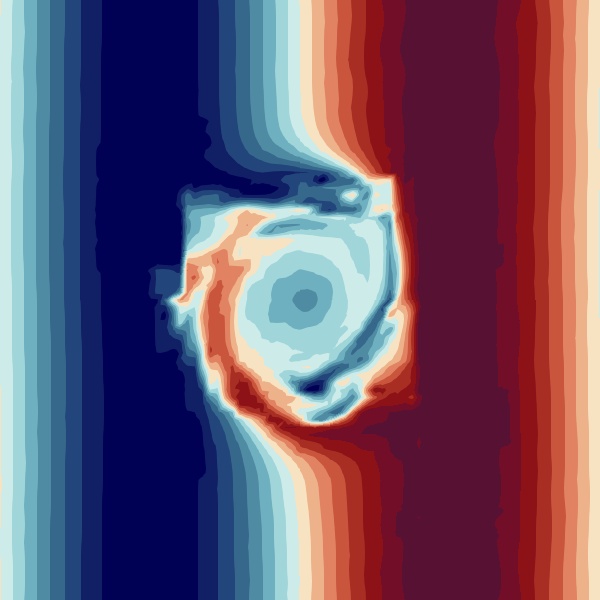}}}
        ~
        \subfloat[$N=40^2$ (without entropy switch)]{\adjustbox{width=0.33\linewidth, valign=b}{\includegraphics{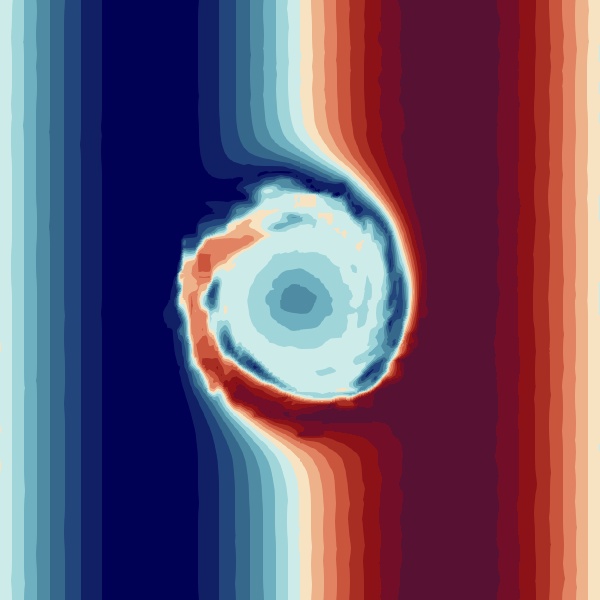}}}
        ~
        \subfloat[$N=80^2$ (without entropy switch)]{\adjustbox{width=0.33\linewidth, valign=b}{\includegraphics{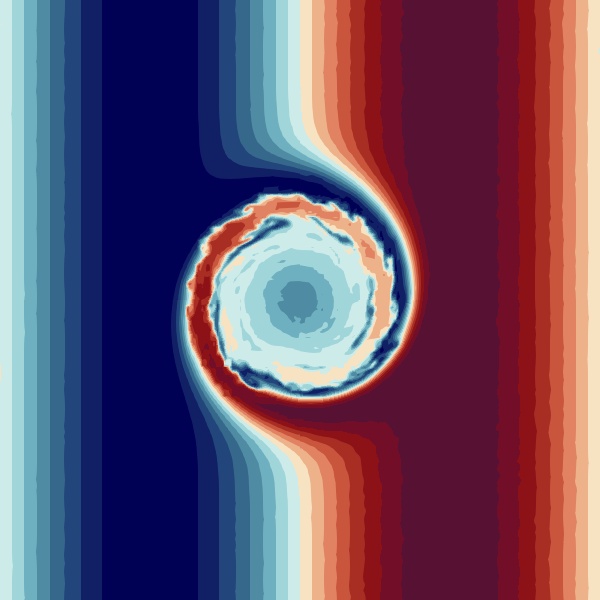}}}
        \caption{\label{fig:icv_swirl}Contours of $\alpha_0\rho_0$ for the multi-species isentropic Euler vortex problem at $t = T = 20$ using a $\mathbb P_4$ approximation with $N = 20^2$ (left), $40^2$ (middle), and $80^2$ (right) elements. Results with and without the entropy switch shown on top and bottom rows, respectively. }
    \end{figure}
    In the multi-species case, forming an exact solution at an arbitrary time is complex. Therefore, we instead use a different procedure to inspect the efficacy of the proposed approach. First, the solution is advanced to some time $t = T$, after which the flow velocities are reversed and then advanced to $t = 2T$. If the initial condition is isentropic, then the solution should be reversible and any irreversibilities stemming from numerical error and dissipation would appear as discrepancies between the solutions at $t = 0$ and $t = 2T$. To yield this isentropic initial condition, we consider the two-species case with $\gamma_0=\gamma_1$, recovered through the specific heat capacities $c_{p,0}=1.4$, $c_{v,0}=1$, $c_{p,1}=2.8$, and $c_{v,1}=2$.

    The contours of the first species density at $t = T = 20$, computed by a $\mathbb P_4$ approximation with varying mesh resolution, are shown in \cref{fig:icv_swirl} with and without the entropy switch. It can evidently be seen that there is a marked difference in the accuracy of the method when the entropy switch is used that is most pronounced in the vicinity of species interfaces. With the switch enabled, very good resolution of the vortex rollup was observed even at $N = 40^2$, showing, at least in the qualitative sense, mesh convergence with respect to the more resolved case of $N=80^2$. However, when the entropy switch was disabled, the enforcement of entropy constraints around species interfaces had a severely detrimental effect on the accuracy of the proposed method, with poor convergence with respect to mesh resolution. 
    
    \begin{figure}[tbhp]
        \centering
        \subfloat[$N=20^2$ (with entropy switch)]{\adjustbox{width=0.33\linewidth, valign=b}{\includegraphics{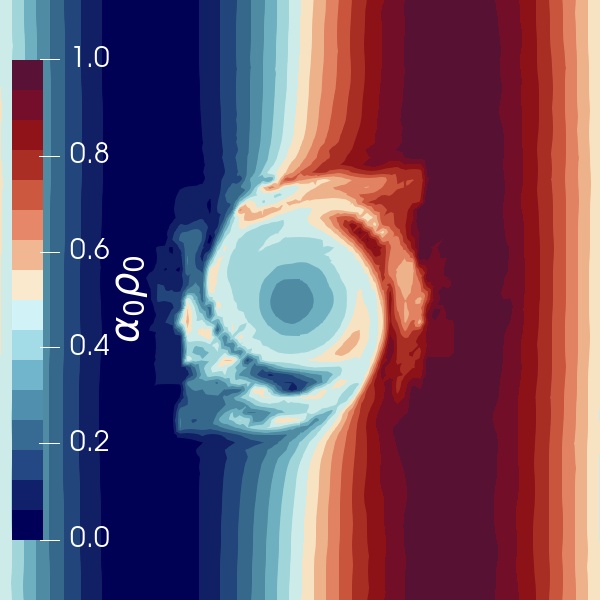}}}
        ~
        \subfloat[$N=40^2$ (with entropy switch)]{\adjustbox{width=0.33\linewidth, valign=b}{\includegraphics{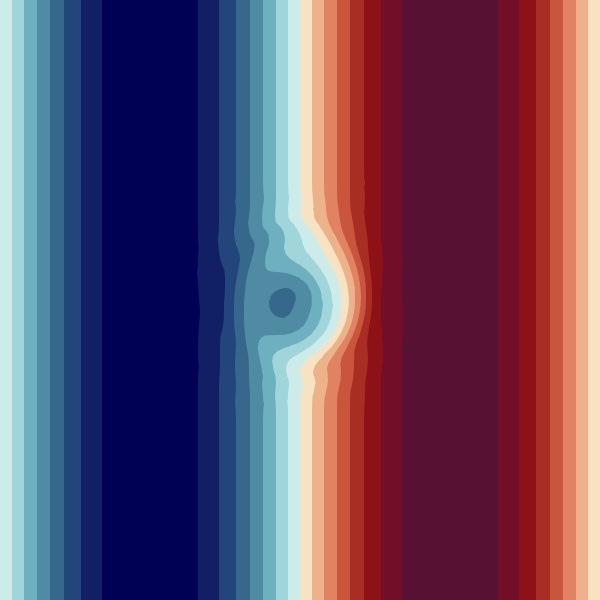}}}
        ~
        \subfloat[$N=80^2$ (with entropy switch)]{\adjustbox{width=0.33\linewidth, valign=b}{\includegraphics{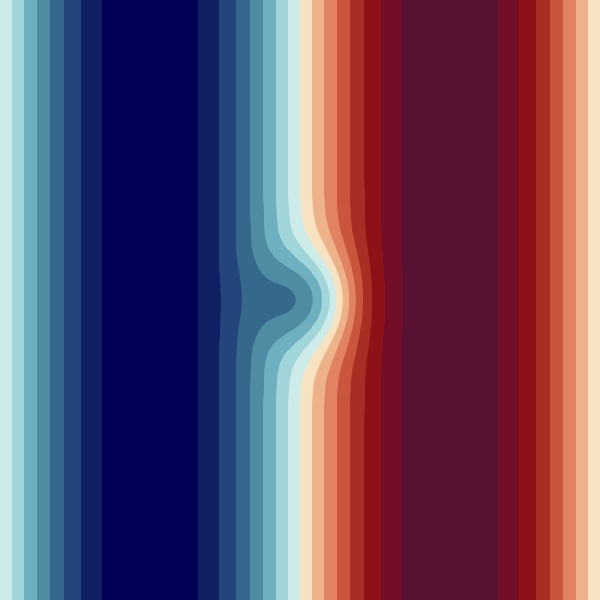}}}\\
        \subfloat[$N=20^2$ (without entropy switch)]{\adjustbox{width=0.33\linewidth, valign=b}{\includegraphics{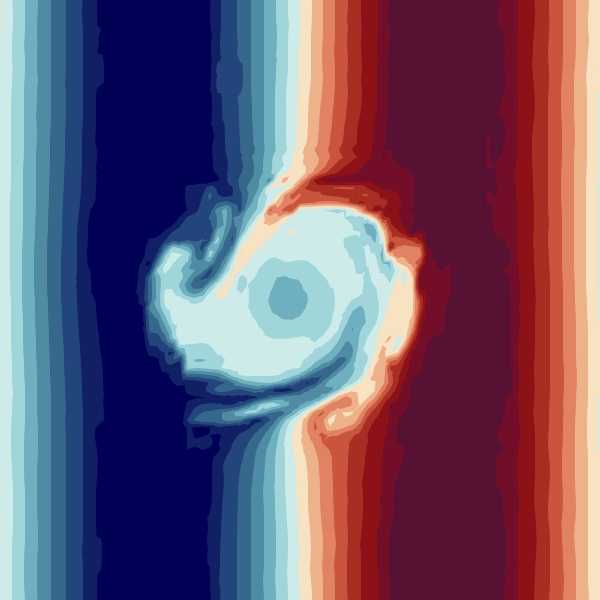}}}
        ~
        \subfloat[$N=40^2$ (without entropy switch)]{\adjustbox{width=0.33\linewidth, valign=b}{\includegraphics{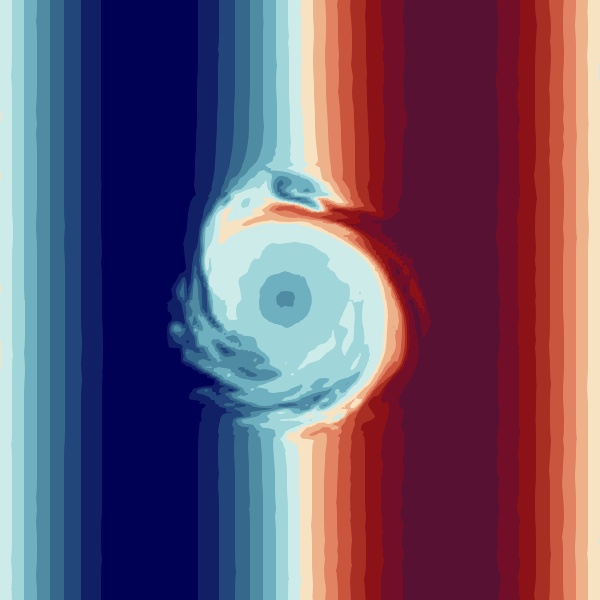}}}
        ~
        \subfloat[$N=80^2$ (without entropy switch)]{\adjustbox{width=0.33\linewidth, valign=b}{\includegraphics{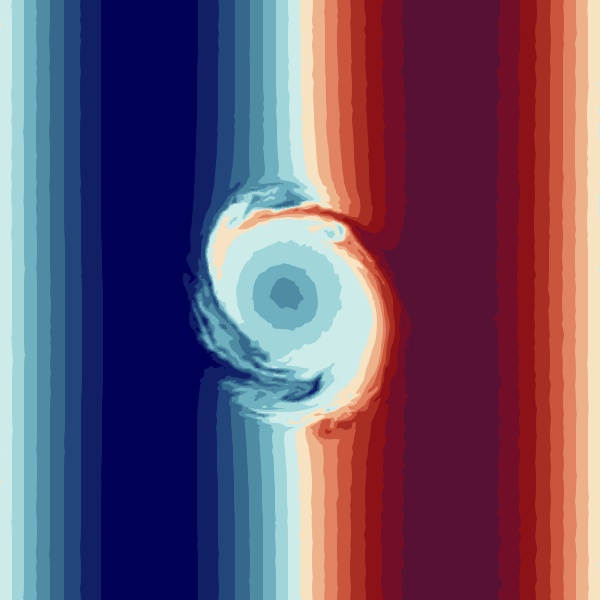}}}
        \caption{\label{fig:icv_swirl2}Contours of $\alpha_0\rho_0$ for the multi-species isentropic Euler vortex problem after flow reversal at $t = 2T = 40$ using a $\mathbb P_4$ approximation with $N = 20^2$ (left), $40^2$ (middle), and $80^2$ (right) elements. Results with and without the entropy switch shown on top and bottom rows, respectively. }
    \end{figure}
    
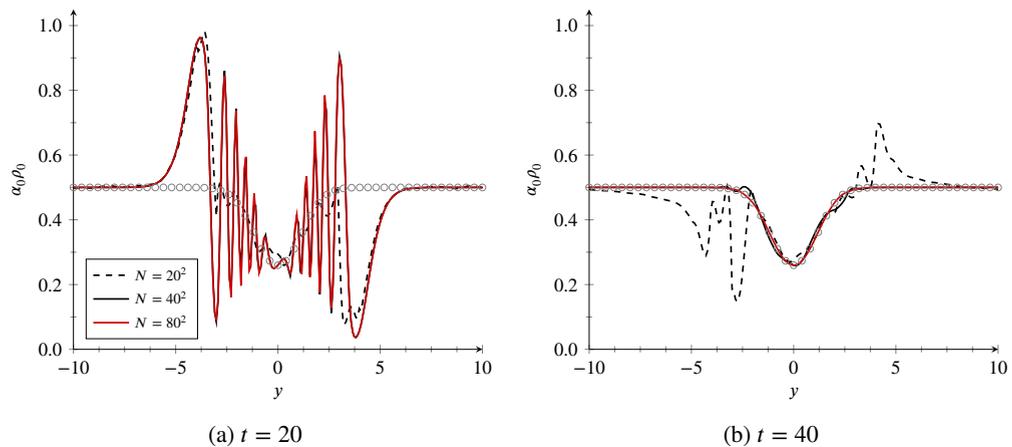
\begin{figure}[tbhp]
        \centering
        \subfloat[$t=20$]{\adjustbox{width=0.4\linewidth, valign=b}{\begin{tikzpicture}[spy using outlines={rectangle, height=3cm,width=2.3cm, magnification=3, connect spies}]
    \begin{axis}
    [
        axis line style={latex-latex},
        axis y line=left,
        axis x line=left,
        xmode=linear,
        ymode=linear,
        xlabel = {$y$},
        ylabel = {$\alpha_0\rho_0$},
        xmin = -10, xmax = 10,
        ymin = 0, ymax = 1.05,
        minor x tick num=3,
        minor y tick num=1,
        ytick = {0,0.2,...,1.0},
        legend cell align={left},
        legend style={font=\scriptsize, at={(0.03, 0.03)},anchor=south west},
        %axis line style={draw=none},
        %tick style={draw=none},
        clip mode=individual,
        x tick label style={/pgf/number format/.cd, precision=1, /tikz/.cd},
        y tick label style={/pgf/number format/.cd, fixed, fixed zerofill, precision=1, /tikz/.cd}
    ]
        % \addplot[color=black!70, style={thin}, only marks, mark=o, mark options={scale=0.6}, mark repeat = 5, mark phase = 0]
        % table[x expr=(\thisrow{x} + 0.5), y=d, col sep=comma]{./figs/data/kq-ref.csv};
        % \addlegendentry{Reference};
        
        \addplot[color=black, style={thick, dashed}] table[x=y, y=a0rh0-nx20-t20-kxrcf, col sep=comma]{./figs/data/vortex-p4-data.csv};
        \addlegendentry{$N=20^2$};

        \addplot[color=black, style={thick}] table[x=y, y=a0rh0-nx40-t20-kxrcf, col sep=comma]{./figs/data/vortex-p4-data.csv};
        \addlegendentry{$N=40^2$};
        
        \addplot[color=red!80!black, style=thick] table[x=y, y=a0rh0-nx80-t20-kxrcf, col sep=comma]{./figs/data/vortex-p4-data.csv};
        \addlegendentry{$N=80^2$};
        
        \addplot[color=black!50, style={thin}, only marks, mark=o, mark options={scale=0.8}, mark repeat = 2, mark phase = 0]
        table[x=y, y=a0rho0, col sep=comma]{./figs/data/vortex_ic.csv};
    \end{axis}
\end{tikzpicture}}}
        ~
        \subfloat[$t=40$]{\adjustbox{width=0.4\linewidth, valign=b}{\begin{tikzpicture}[spy using outlines={rectangle, height=3cm,width=2.3cm, magnification=3, connect spies}]
    \begin{axis}
    [
        axis line style={latex-latex},
        axis y line=left,
        axis x line=left,
        clip mode=individual,
        xmode=linear,
        ymode=linear,
        xlabel = {$y$},
        ylabel = {$\alpha_0\rho_0$},
        xmin = -10, xmax = 10,
        ymin = 0, ymax = 1.05,
        minor x tick num=3,
        minor y tick num=1,
        ytick = {0,0.2,...,1.0},
        legend cell align={left},
        legend style={font=\scriptsize, at={(0.03, 0.03)},anchor=south west},
        %axis line style={draw=none},
        %tick style={draw=none},
        clip mode=individual,
        x tick label style={/pgf/number format/.cd, precision=1, /tikz/.cd},
        y tick label style={/pgf/number format/.cd, fixed, fixed zerofill, precision=1, /tikz/.cd}
    ]
        % \addplot[color=black!70, style={thin}, only marks, mark=o, mark options={scale=0.6}, mark repeat = 5, mark phase = 0]
        % table[x expr=(\thisrow{x} + 0.5), y=d, col sep=comma]{./figs/data/kq-ref.csv};
        % \addlegendentry{Reference};
        
        \addplot[color=black, style={thick, dashed}] table[x=y, y=a0rh0-nx20-t40-kxrcf, col sep=comma]{./figs/data/vortex-p4-data.csv};
        % \addlegendentry{$N=20$};

        \addplot[color=black, style={thick}] table[x=y, y=a0rh0-nx40-t40-kxrcf, col sep=comma]{./figs/data/vortex-p4-data.csv};
        % \addlegendentry{$N=40$};
        
        \addplot[color=red!80!black, style=thick] table[x=y, y=a0rh0-nx80-t40-kxrcf, col sep=comma]{./figs/data/vortex-p4-data.csv};
        % \addlegendentry{$N=80$};

        \addplot[color=black!50, style={thin}, only marks, mark=o, mark options={scale=0.8}, mark repeat = 2, mark phase = 0]
        table[x=y, y=a0rho0, col sep=comma]{./figs/data/vortex_ic.csv};
        
    \end{axis}
\end{tikzpicture}}}
        \caption{\label{fig:icv_slice}Cross-section of $\alpha_0\rho_0$ at $x=0$ for the multi-species isentropic Euler vortex problem at $t = T = 20$ (left) and after flow reversal at $t = 2T = 40$ (right) using a $\mathbb P_4$ approximation with varying mesh resolution computed \textit{with} the entropy switch. Initial $\alpha_0\rho_0$ field at $t = 0$ shown by grey markers.}
    \end{figure}

\begin{figure}[tbhp]
        \centering
        \subfloat[$t=20$]{\adjustbox{width=0.4\linewidth, valign=b}{\begin{tikzpicture}[spy using outlines={rectangle, height=3cm,width=2.3cm, magnification=3, connect spies}]
    \begin{axis}
    [
        axis line style={latex-latex},
        axis y line=left,
        axis x line=left,
        xmode=linear,
        ymode=linear,
        xlabel = {$y$},
        ylabel = {$\alpha_0\rho_0$},
        xmin = -10, xmax = 10,
        ymin = 0, ymax = 1.05,
        minor x tick num=3,
        minor y tick num=1,
        ytick = {0,0.2,...,1.0},
        legend cell align={left},
        legend style={font=\scriptsize, at={(0.03, 0.03)},anchor=south west},
        %axis line style={draw=none},
        %tick style={draw=none},
        clip mode=individual,
        x tick label style={/pgf/number format/.cd, precision=1, /tikz/.cd},
        y tick label style={/pgf/number format/.cd, fixed, fixed zerofill, precision=1, /tikz/.cd}
    ]
        
        \addplot[color=black, style=thick] table[x=y, y=a0rh0-nx80-t20-nokxrcf, col sep=comma]{./figs/data/vortex-p4-data.csv};
        \addlegendentry{Without entropy switch};
        
        \addplot[color=red!80!black, style=thick] table[x=y, y=a0rh0-nx80-t20-kxrcf, col sep=comma]{./figs/data/vortex-p4-data.csv};
        \addlegendentry{With entropy switch};
        
        \addplot[color=black!50, style={thin}, only marks, mark=o, mark options={scale=0.8}, mark repeat = 2, mark phase = 0]
        table[x=y, y=a0rho0, col sep=comma]{./figs/data/vortex_ic.csv};
    \end{axis}
\end{tikzpicture}}}
        ~
        \subfloat[$t=40$]{\adjustbox{width=0.4\linewidth, valign=b}{\begin{tikzpicture}[spy using outlines={rectangle, height=3cm,width=2.3cm, magnification=3, connect spies}]
    \begin{axis}
    [
        axis line style={latex-latex},
        axis y line=left,
        axis x line=left,
        xmode=linear,
        ymode=linear,
        xlabel = {$y$},
        ylabel = {$\alpha_0\rho_0$},
        xmin = -10, xmax = 10,
        ymin = 0, ymax = 1.05,
        minor x tick num=3,
        minor y tick num=1,
        ytick = {0,0.2,...,1.0},
        legend cell align={left},
        legend style={font=\scriptsize, at={(0.03, 0.03)},anchor=south west},
        %axis line style={draw=none},
        %tick style={draw=none},
        clip mode=individual,
        x tick label style={/pgf/number format/.cd, precision=1, /tikz/.cd},
        y tick label style={/pgf/number format/.cd, fixed, fixed zerofill, precision=1, /tikz/.cd}
    ]
        
        \addplot[color=black, style=thick] table[x=y, y=a0rh0-nx80-t40-nokxrcf, col sep=comma]{./figs/data/vortex-p4-data.csv};
        % \addlegendentry{Without KXRCF sensor};
        
        \addplot[color=red!80!black, style=thick] table[x=y, y=a0rh0-nx80-t40-kxrcf, col sep=comma]{./figs/data/vortex-p4-data.csv};
        % \addlegendentry{With KXRCF sensor};
        
        \addplot[color=black!50, style={thin}, only marks, mark=o, mark options={scale=0.8}, mark repeat = 2, mark phase = 0]
        table[x=y, y=a0rho0, col sep=comma]{./figs/data/vortex_ic.csv};
    \end{axis}
\end{tikzpicture}}}
        \caption{\label{fig:icv_slice_comp}Comparison of the cross-section of $\alpha_0\rho_0$ at $x=0$ for the multi-species isentropic Euler vortex problem at $t = T = 20$ (left) and after flow reversal at $t = 2T = 40$ (right) using a $\mathbb P_4$ approximation with $N = 80^2$ computed with the entropy switch (red) and without the entropy switch (black). Initial $\alpha_0\rho_0$ field at $t = 0$ shown by grey markers.}
    \end{figure}
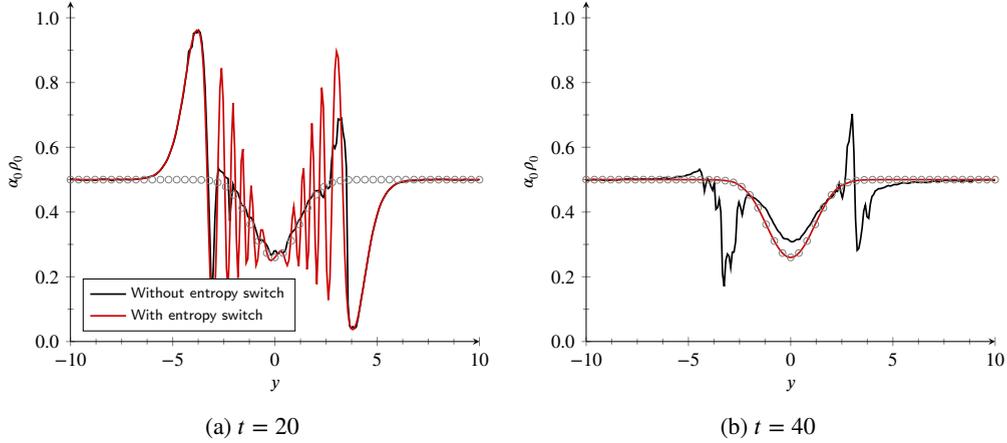

    After flow reversal at $t = 2T$, shown in \cref{fig:icv_swirl2}, the difference between the accuracy of the method with and without the entropy switch became even more pronounced. While the approach with the entropy switch was able to recover the initial conditions reasonably well with $N=40^2$ and very well with $N = 80^2$, the approach without the entropy switch showed noticeable numerical artefacts at all resolutions as a result of the excessive numerical dissipation introduced by the entropy constraints at species interfaces. As a result, the predictions at $N=80^2$ without the entropy switch were on the order of the much less resolved $N=20^2$ simulation that did utilise the entropy switch. These observations are highlighted in \cref{fig:icv_slice} which shows a cross-section of the first species density at $x=0$ at both the flow reversal time and final time computed with the entropy switch. It can be seen that the approach can resolve the highly-oscillatory nature of the vortex rollup and recover the initial conditions very well. Furthermore, a comparison of the results with and without the entropy switch is shown on the same cross-section in \cref{fig:icv_slice_comp}. The substandard resolving ability for the vortex rollup and the flow irreversibilities introduced by the significant numerical dissipation of the approach without the entropy switch can be clearly observed. These results showcase both the resolving ability of the scheme and the importance of adapting the entropy bounds by the proposed approach. 
    
\subsubsection{Quirk and Karni shock tube}
    To evaluate the proposed scheme for more complex flow physics including shock waves, contact discontinuities, and rarefaction waves, the one-dimensional shock tube problem of \citet{Quirk1996}, a simplification of a shock-bubble interaction, was considered. In this test case, the various features of the Riemann problem are observed as well as discontinuities between species, which can introduce pressure oscillations in conservative approaches where pressure equilibrium is not guaranteed. 
    The problem is solved in a quasi-one-dimensional form on the domain $\Omega = [0,1]\times[-0.05, 0.05]$, and the initial conditions are given by \cref{tab:quirk_states}.

    \begin{table}[tbhp]
        \centering
        \caption{\label{tab:quirk_states} Initial conditions for the Quirk and Karni shock tube problem.}
        \begin{tabular}{cccccccc}
            \toprule
            Region & Extent & $\rho_0$ & $\rho_1$ & $u$ & $P$ & $\gamma$ & $c_v$ \\ \midrule
            L & $x\leq0.25$& 1.3765 & 0 & 0.3948 & 1.57 & 1.4 & 0.72 \\
            R & $0.25<x\leq0.4$, $x > 0.6$ & 1 & 0 & 0 & 1 & 1.4 & 0.72 \\
            B & $0.4<x\leq0.6$ & 0 & 0.138 & 0 & 1 & 1.67 & 2.52\\
            \bottomrule
        \end{tabular}
    \end{table}

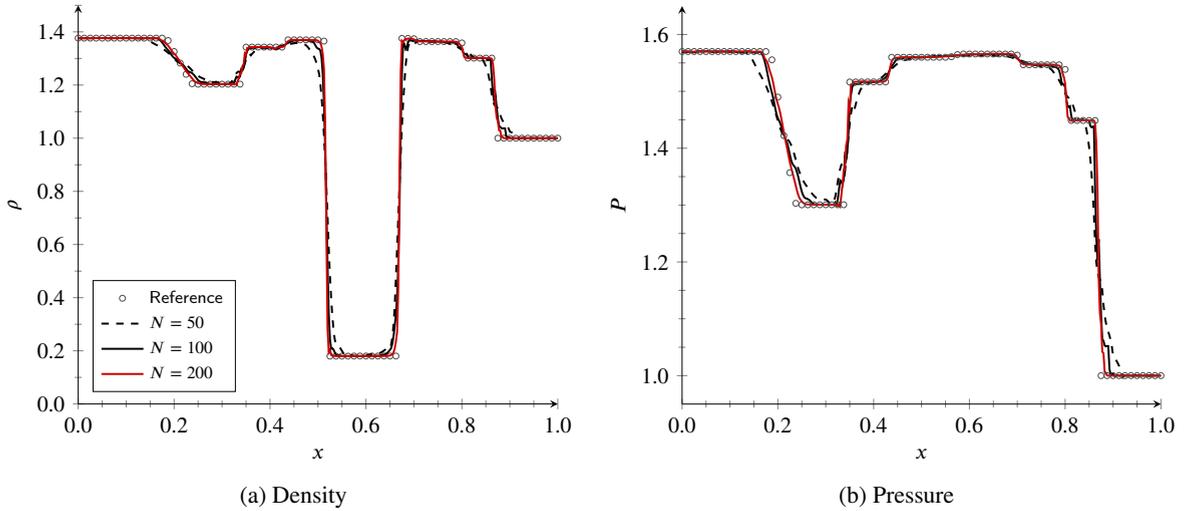
\begin{figure}[tbhp]
        \centering
        \subfloat[Density]{\label{fig:kq_p3_density}\adjustbox{width=0.47\linewidth, valign=b}{\begin{tikzpicture}[spy using outlines={rectangle, height=3cm,width=2.3cm, magnification=3, connect spies}]
    \begin{axis}
    [
        axis line style={latex-latex},
        axis y line=left,
        axis x line=left,
        xmode=linear,
        ymode=linear,
        xlabel = {$x$},
        ylabel = {$\rho$},
        xmin = 0, xmax = 1,
        ymin = 0, ymax = 1.5,
        minor x tick num=3,
        minor y tick num=1,
        ytick = {0,0.2,...,1.5},
        legend cell align={left},
        legend style={font=\scriptsize, at={(0.03, 0.03)},anchor=south west},
        %axis line style={draw=none},
        %tick style={draw=none},
        clip mode=individual,
        x tick label style={/pgf/number format/.cd, fixed, fixed zerofill, precision=1, /tikz/.cd},
        y tick label style={/pgf/number format/.cd, fixed, fixed zerofill, precision=1, /tikz/.cd}
    ]
        \addplot[color=black!70, style={thin}, only marks, mark=o, mark options={scale=0.6}, mark repeat = 5, mark phase = 0]
        table[x expr=(\thisrow{x} + 0.5), y=d, col sep=comma]{./figs/data/kq-ref.csv};
        \addlegendentry{Reference};
        
        \addplot[color=black, style={thick, dashed}] table[x expr=(\thisrow{x} + 0.5), y=d-nx50, col sep=comma]{./figs/data/kq-p3-kxrcf-new_p1.csv};
        \addlegendentry{$N=50$};

        \addplot[color=black, style={thick}] table[x expr=(\thisrow{x} + 0.5), y=d-nx100, col sep=comma]{./figs/data/kq-p3-kxrcf-new_p1.csv};
        \addlegendentry{$N=100$};
        
        \addplot[color=red!80!black, style={thick}] table[x expr=(\thisrow{x} + 0.5), y=d-nx200, col sep=comma]{./figs/data/kq-p3-kxrcf-new_p1.csv};
        \addlegendentry{$N=200$};
        
    \end{axis}
\end{tikzpicture}}}
        ~
        \subfloat[Pressure]{\label{fig:kq_p3_pressure}\adjustbox{width=0.47\linewidth}{\begin{tikzpicture}[spy using outlines={rectangle, height=3cm,width=2.3cm, magnification=3, connect spies}]
    \begin{axis}
    [
        axis line style={latex-latex},
        axis y line=left,
        axis x line=left,
        xmode=linear,
        ymode=linear,
        xlabel = {$x$},
        ylabel = {$P$},
        xmin = 0, xmax = 1,
        ymin = 0.95, ymax = 1.65,
        ymin = 0.95, ymax = 1.65,
        minor x tick num=3,
        minor y tick num=3,
        ytick = {1,1.2,...,1.6},
        legend cell align={left},
        legend style={font=\scriptsize, at={(0.97, 0.97)},anchor=north east},
        %axis line style={draw=none},
        %tick style={draw=none},
        x tick label style={/pgf/number format/.cd, fixed, fixed zerofill, precision=1, /tikz/.cd},
        y tick label style={/pgf/number format/.cd, fixed, fixed zerofill, precision=1, /tikz/.cd},
    ]
        \addplot[color=black!70, style={thin}, only marks, mark=o, mark options={scale=0.6}, mark repeat = 5, mark phase = 0]
        table[x expr=(\thisrow{x} + 0.5), y=p, col sep=comma]{./figs/data/kq-ref.csv};
        
        \addplot[color=black, style={thick, dashed}] table[x expr=(\thisrow{x} + 0.5), y=p-nx50, col sep=comma]{./figs/data/kq-p3-kxrcf-new_p1.csv};

        \addplot[color=black, style={thick}] table[x expr=(\thisrow{x} + 0.5), y=p-nx100, col sep=comma]{./figs/data/kq-p3-kxrcf-new_p1.csv};
        
        \addplot[color=red!80!black, style={thick}] table[x expr=(\thisrow{x} + 0.5), y=p-nx200, col sep=comma]{./figs/data/kq-p3-kxrcf-new_p1.csv};
        
    \end{axis}
\end{tikzpicture}}}
        \caption{\label{fig:kq_p3}Profiles of density (left) and pressure (right) for the Quirk and Karni shock tube problem at $t = 0.35$ computed using a $\mathbb{P}_3$ approximation with varying mesh resolution.}
    \end{figure}

    For this experiment, a $\mathbb P_3$ approximation was used with varying mesh resolution. The resulting density and pressure fields at $t = 0.35$ are shown in \cref{fig:kq_p3}. A reference solution was computed with a highly resolved, first-order Godunov \citep{Godunov1959} scheme with $N = 5{\cdot}10^4$ elements. It can be seen that the proposed approach shows good convergence to the reference solution with increasing mesh resolution, with excellent agreement at $N = 200$. Furthermore, better prediction of discontinuities in the flow was observed with increasing resolution, with less numerical dissipation introduced by scheme. In the pressure field, the oscillations associated with the pressure equilibrium problem \citep{Abgrall1996}, commonly seen in conservative approaches in the region $0.4<x<0.7$, were not visually noticeable. It is expected that these oscillations reduce with increasing mesh resolution~\citep{Abgrall1996,Johnson2020}, such that their effects are minimal for well-resolved simulations. 

\subsubsection{Helium-air shock-bubble interaction}
 
    \begin{figure}[tbhp]
        \centering
        \adjustbox{width=0.7\linewidth, valign=b}{\begin{tikzpicture}[scale=1.2]
    \draw[black, thick] (0, -1) rectangle (10, 1);

    \draw[black, thick] (8,0) circle (50/89);

    \draw[black, thick, dashed] (9,-1) -- (9,1);
    \draw[black, thin] (8,-1.04) -- (8,-0.96);
    \draw[black, thin] (8,1.04) -- (8,0.96);
    \draw[black, thin] (-0.04,0) -- (0.04,0);
    \draw[black, thin] (9.96,0) -- (10.04,0);

    \draw[black, thin, dashed] (8, 50/89) -- (6.5, 50/89);
    \draw[black, thin, dashed] (8, -50/89) -- (6.5, -50/89);
    \draw[black, latex-latex] (6.6, -50/89) -- (6.6, 50/89) node [midway, xshift=-2em] {$10/89$};

    \node[text=black] at (3, 0) {
    \begin{tabular}{r} $\rho_l$ \\ $P_l$ \\ $\gamma_a$ \end{tabular}};

    \node[text=black] at (8, 0) {
    \begin{tabular}{r} $\rho_b$ \\ $P_b$ \\ $\gamma_b$ \end{tabular}};

    \node[text=black] at (9.5, 0) {
    \begin{tabular}{r} $\rho_r$ \\ $P_r$ \\ $\gamma_a$ \end{tabular}};
    
    \draw[black, -latex] (-0.2, -1.15) -- (0.6, -1.15);
    \node[text=black] at (0.725, -1.15) {$x$};
    \draw[black, -latex] (-0.15, -1.2) -- (-0.15, -0.4);
    \node[text=black] at (-0.15, -0.25) {$y$};

    \node[text=black] at (0, -1.4) {$0$};
    \node[text=black] at (8, -1.4) {$0.8$};
    \node[text=black] at (9, -1.4) {$0.9$};
    \node[text=black] at (10, -1.4) {$1$};
    
    \node[text=black] at (10.5, -1) {$-0.1$};
    \node[text=black] at (10.5, 0) {$0$};
    \node[text=black] at (10.5, 1) {$0.1$};
    
\end{tikzpicture}}
        \caption{\label{fig:bubble_ic}Schematic of the domain and initial conditions for the helium-air shock-bubble interaction problem.}
    \end{figure}
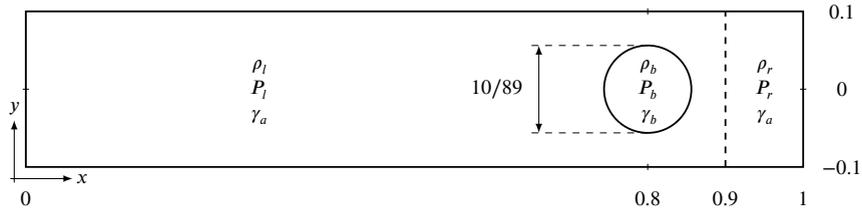
The extension to more complex flow physics and unstructured meshes was performed through the helium-air shock-bubble interaction problem, a well-studied case first experimentally investigated by \citet{Haas1987} and numerically simulated by \citet{Quirk1996}. The interaction between the travelling shock wave and the helium gas bubble seeds Richtmyer--Meshkov instabilities at the interface, yielding complex vortical flow. Accurately predicting this interaction is the primary challenge in this test case as stabilisation methods with excessive numerical dissipation tend to smear the interface and associated small-scale flow features.
    
    \begin{table}[tbhp]
        \centering
        \caption{\label{tab:bubble_states} Initial conditions for the helium-air shock-bubble interaction problem.}
        \begin{tabular}{cccccccc}
            \toprule
            Region & $\rho_0$ & $\rho_1$ & $u$ & $v$ & $P$ & $\gamma$ & $c_v$ \\ \midrule
            L & 1 & 0 & 0 & 0 & 1 & 1.3986 & 0.72 \\
            R & 1.37636 & 0 & -0.55957 & 0 & 1.5698 & 1.3986 & 0.72 \\
            B & 0 & 0.18187 & 0 & 0 & 1  & 1.6467 & 2.44\\
            \bottomrule
        \end{tabular}
    \end{table}

The numerical setup consists of a bubble of radius $R = 5/89$ filled with helium contaminated with 28\% air by mass. This bubble is centred at $x = 0.8$, $y = 0$ within a channel of unit length and height of $0.2$ filled with air. The resulting numerical domain is set as $\Omega = [0, 1] \times [-0.1, 0.1]$, shown in \cref{fig:bubble_ic}. At $x = 0.9$, a normal shock corresponding to a Mach number of $M = 1.22$ is imposed. The flow is normalised to unit density and pressure by the post-shock air state, denoted by the subscript $l$. The pre-shock air state, denoted by the subscript $r$, is computed via the Rankine--Hugoniot conditions. The bubble state, denoted with the subscript $b$, is set to be in thermal and mechanical equilibrium with the surrounding air. The initial conditions as well as the necessary gas constants for the three states are presented in \cref{tab:bubble_states}. 

    \begin{figure}[tbhp]
        \centering
        \subfloat[Coarse mesh ($h/R = 50$)]{\adjustbox{width=0.33\linewidth, valign=b}{\includegraphics{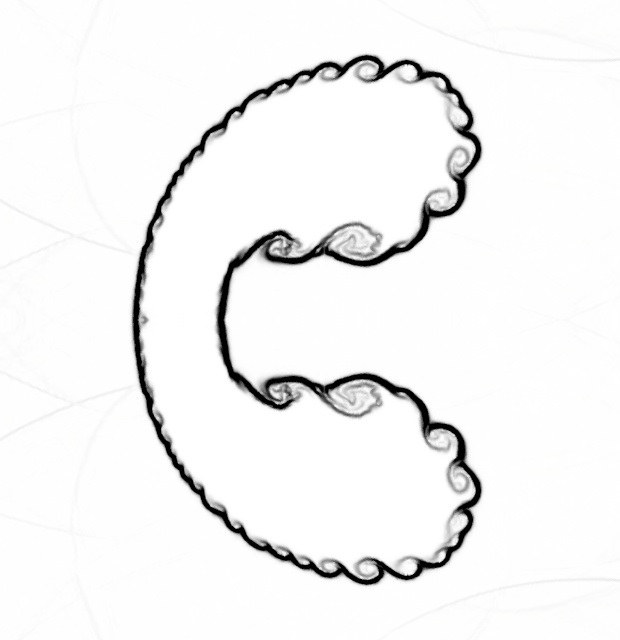}}}
        ~
        \subfloat[Medium mesh ($h/R = 100$)]{\adjustbox{width=0.33\linewidth, valign=b}{\includegraphics{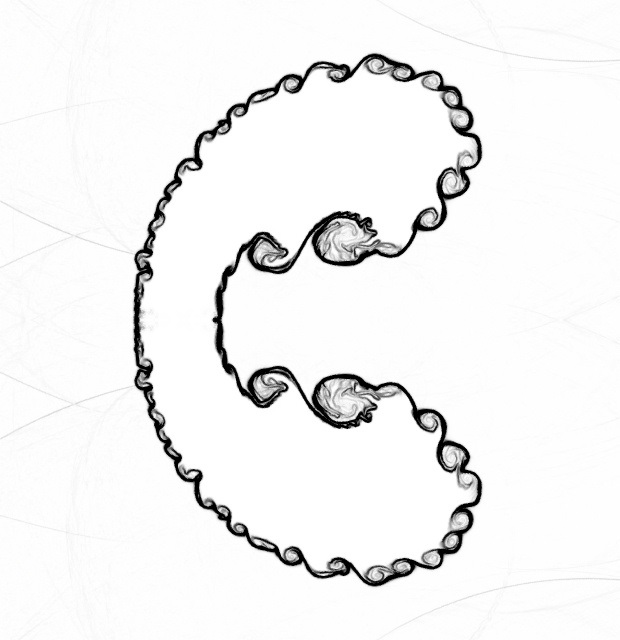}}}
        ~
        \subfloat[Fine mesh ($h/R = 200$)]{\adjustbox{width=0.33\linewidth, valign=b}{\includegraphics{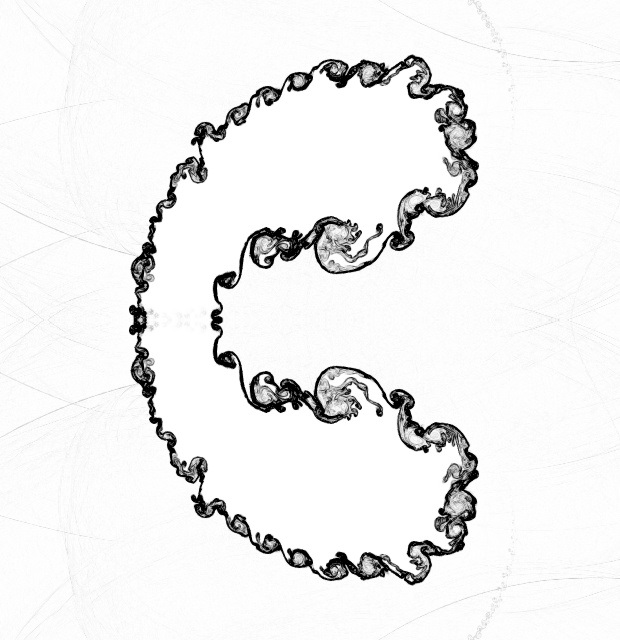}}}
        ~
        \caption{\label{fig:shockbubble_t25} Numerical Schlieren diagrams for the helium-air shock-bubble interaction at $t = 0.25$ computed on a coarse (left), medium (middle), and fine (right) unstructured triangular mesh with a $\mathbb P_3$ approximation. Solution on the half-domain is reflected across the $x$-axis for visualisation.}
    \end{figure}
    
    \begin{figure}[tbhp]
        \centering
        \subfloat[Coarse mesh ($h/R = 50$)]{\adjustbox{width=0.33\linewidth, valign=b}{\includegraphics{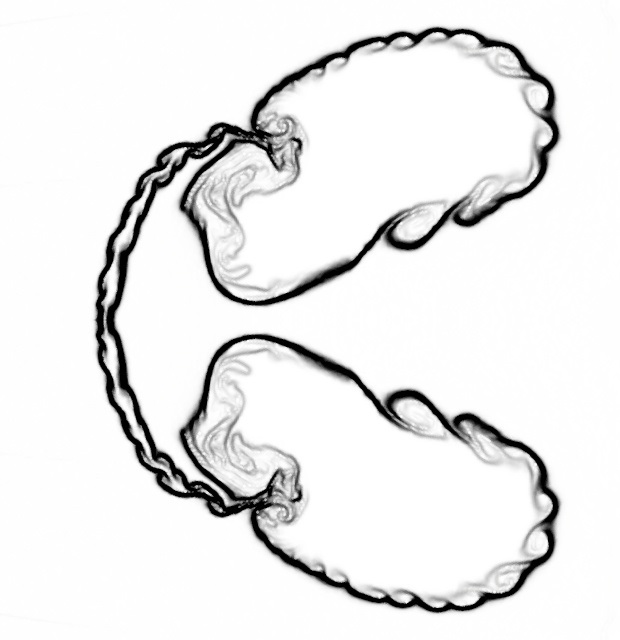}}}
        ~
        \subfloat[Medium mesh ($h/R = 100$)]{\adjustbox{width=0.33\linewidth, valign=b}{\includegraphics{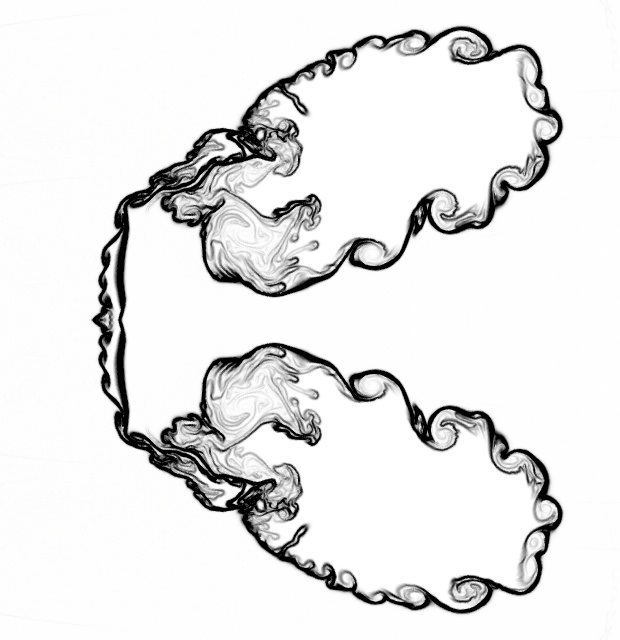}}}
        ~
        \subfloat[Fine mesh ($h/R = 200$)]{\adjustbox{width=0.33\linewidth, valign=b}{\includegraphics{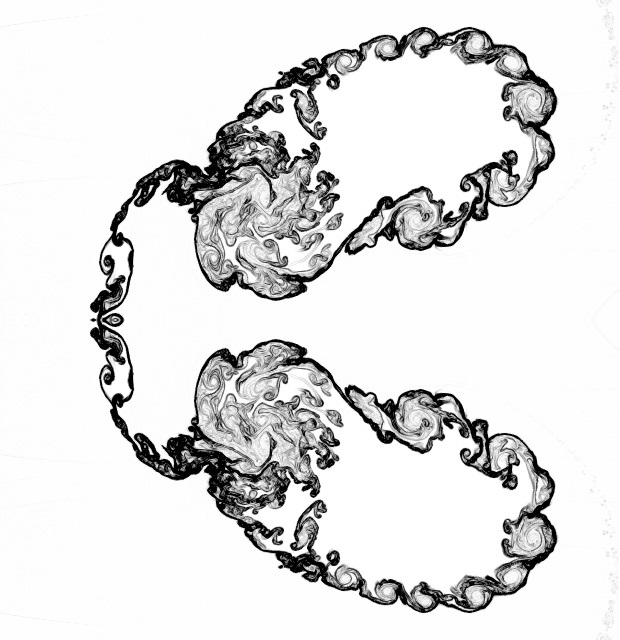}}}
        ~
        \caption{\label{fig:shockbubble_t39} Numerical Schlieren diagrams for the helium-air shock-bubble interaction at $t = 0.39$ computed on a coarse (left), medium (middle), and fine (right) unstructured triangular mesh with a $\mathbb P_3$ approximation. Solution on the half-domain is reflected across the $x$-axis for visualisation.}
    \end{figure}
    
The problem was solved using a $\mathbb P_3$ approximation on unstructured meshes. Due to the symmetry of the problem, only half of the domain ($y > 0$) was solved for, with symmetry (slip adiabatic wall) boundary conditions imposed along the lower wall. Similarly, slip adiabatic wall boundary conditions were used for the upper wall. Dirichlet boundary conditions were used for the inlet ($x = 1$) and Neumann boundary conditions for the outlet ($x = 0$). A series of unstructured triangular meshes were generated, with the coarse, medium, and fine meshes corresponding to average edge lengths of $h/R = 50$, $100$, and $200$, respectively. The resulting numerical Schlieren diagrams at $t = 0.25$, showing the magnitude of the density gradient, are shown in \cref{fig:shockbubble_t25} for the various mesh resolutions. The rollup of vortices along the bubble interface were clearly observed, with smaller scale flow features appearing with increasing resolution. At a later time of $t = 0.39$, shown in \cref{fig:shockbubble_t39}, the bubble showed more lateral spread and distinct vortical structures, and similar observations with respect to the mesh resolution were drawn. 

    \begin{figure}[tbhp]
        \centering
        \subfloat[$t = 0.25$ ($426$ $\mu$s)]{\adjustbox{width=0.33\linewidth, valign=b}{\includegraphics{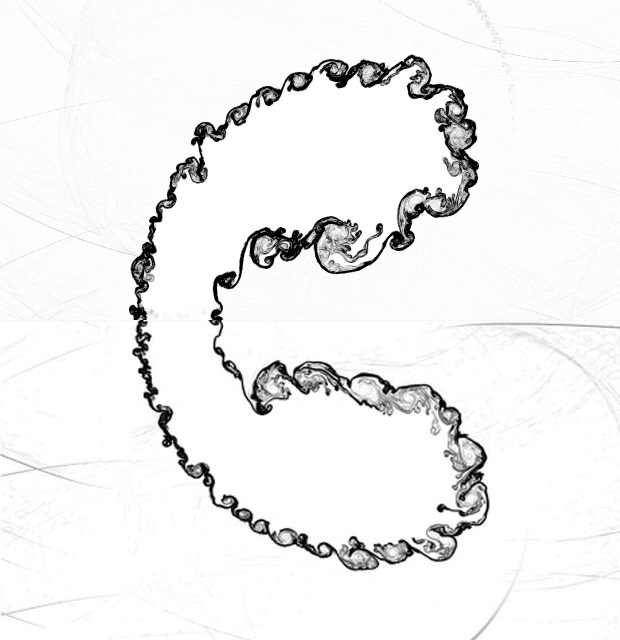}}}
        ~
        \subfloat[$t = 0.39$ ($674$ $\mu$s)]{\adjustbox{width=0.33\linewidth, valign=b}{\includegraphics{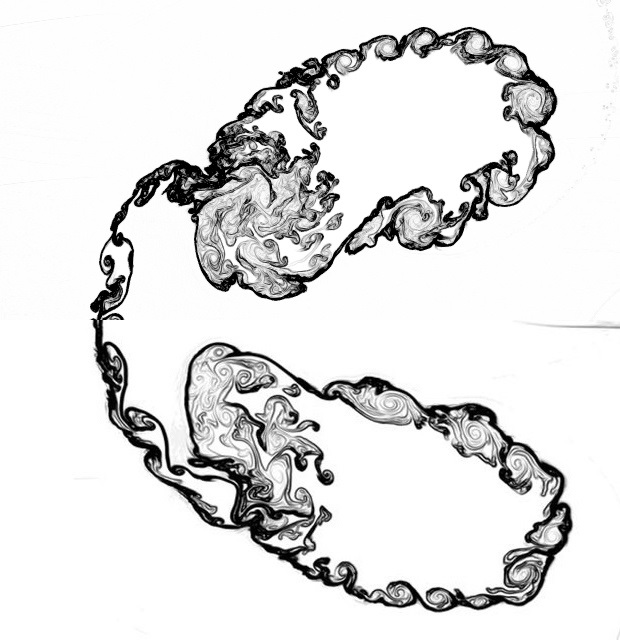}}}
        ~
        \caption{\label{fig:shockbubble_comp} Comparison of the numerical Schlieren diagrams for the helium-air shock-bubble interaction between the present work (top) and \citet{Kundu2021} (bottom) at $t = 0.25$ (left) and $t = 0.39$ (right). Corresponding simulation time of \citet{Kundu2021} shown in parentheses.}
    \end{figure}
    
A comparison of the proposed approach on the fine mesh and the results of \citet{Kundu2021}, computed with a ninth-order upwind finite difference scheme, is shown in \cref{fig:shockbubble_comp} at the two simulation times. Between the two methods, the effective mesh resolution in terms of solution point spacing was roughly similar. Very good agreement between the two approaches was observed, with near identical predictions of the dominant flow structures in the problem as well as good agreement in the shape and size of the small-scale flow features. 

\subsubsection{Triple-point shock interaction}
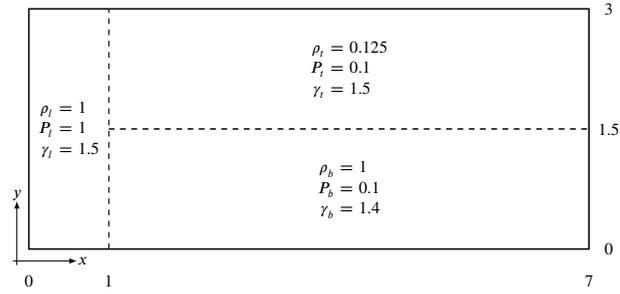
\begin{figure}[tbhp]
        \centering
        \adjustbox{width=0.5\linewidth, valign=b}{\begin{tikzpicture}[scale=1.5]
    \draw[black, thick] (0, 0) rectangle (7, 3);

    \draw[black, dashed] (1, 0) -- (1, 3);
    \draw[black, dashed] (1, 1.5) -- (7, 1.5);
    \draw[black, dashed] (1, 1.5) -- (7, 1.5);
    
    \draw[black, -latex] (-0.2, -0.15) -- (0.6, -0.15);
    \node[text=black] at (0.675, -0.15) {$x$};
    \draw[black, -latex] (-0.15, -0.2) -- (-0.15, 0.6);
    \node[text=black] at (-0.15, 0.675) {$y$};
    
    \node[text=black] at (0, -0.4) {$0$};
    \node[text=black] at (1, -0.4) {$1$};
    \node[text=black] at (7, -0.4) {$7$};
    
    \node[text=black] at (7.25, 0) {$0$};
    \node[text=black] at (7.25, 1.5) {$1.5$};
    \node[text=black] at (7.25, 3) {$3$};

    \node[text=black] at (0.5, 1.5) {
    \begin{tabular}{r@{ = }l} $\rho_l$ & $1$ \\ $P_l$ & $1$ \\ $\gamma_l$ & $1.5$ \end{tabular}};

    \node[text=black] at (4, 2.25) {
    \begin{tabular}{r@{ = }l} $\rho_t$ & $0.125$ \\ $P_t$ & $0.1$ \\ $\gamma_t$ & $1.5$ \end{tabular}};
    
    \node[text=black] at (4, 0.75) {
    \begin{tabular}{r@{ = }l} $\rho_b$ & $1$ \\ $P_b$ & $0.1$ \\ $\gamma_b$ & $1.4$ \end{tabular}};
    
\end{tikzpicture}}
        \caption{\label{fig:trip_pnt}Schematic of the domain and initial conditions for the triple-point shock interaction problem.}
    \end{figure}

As a final inviscid test case, the triple-point shock interaction problem of \citet{Galera2010} was used to evaluate the method for complex shock-vortex interactions. The problem consists of a two-dimensional, two-species Riemann problem, with an overpressure state driving the rollup of a contact discontinuity between species. The domain is set as $\Omega = [0, 7] \times [0, 3]$ with slip adiabatic walls on all four sides, and the initial conditions for the three states are shown in \cref{fig:trip_pnt}. The initial velocity is set to zero throughout the domain. To achieve the desired specific heat ratios of $\gamma_0 = \gamma_l = \gamma_t = 1.5$ and $\gamma_1 = \gamma_b = 1.4$, we simply set $c_{v,0} = c_{v,1} = 1$ and $c_p = \gamma$, respectively. 

\begin{figure}[tbhp]
        \centering
        \subfloat[$t = 1$]{\adjustbox{width=0.47\linewidth, valign=b}{\includegraphics{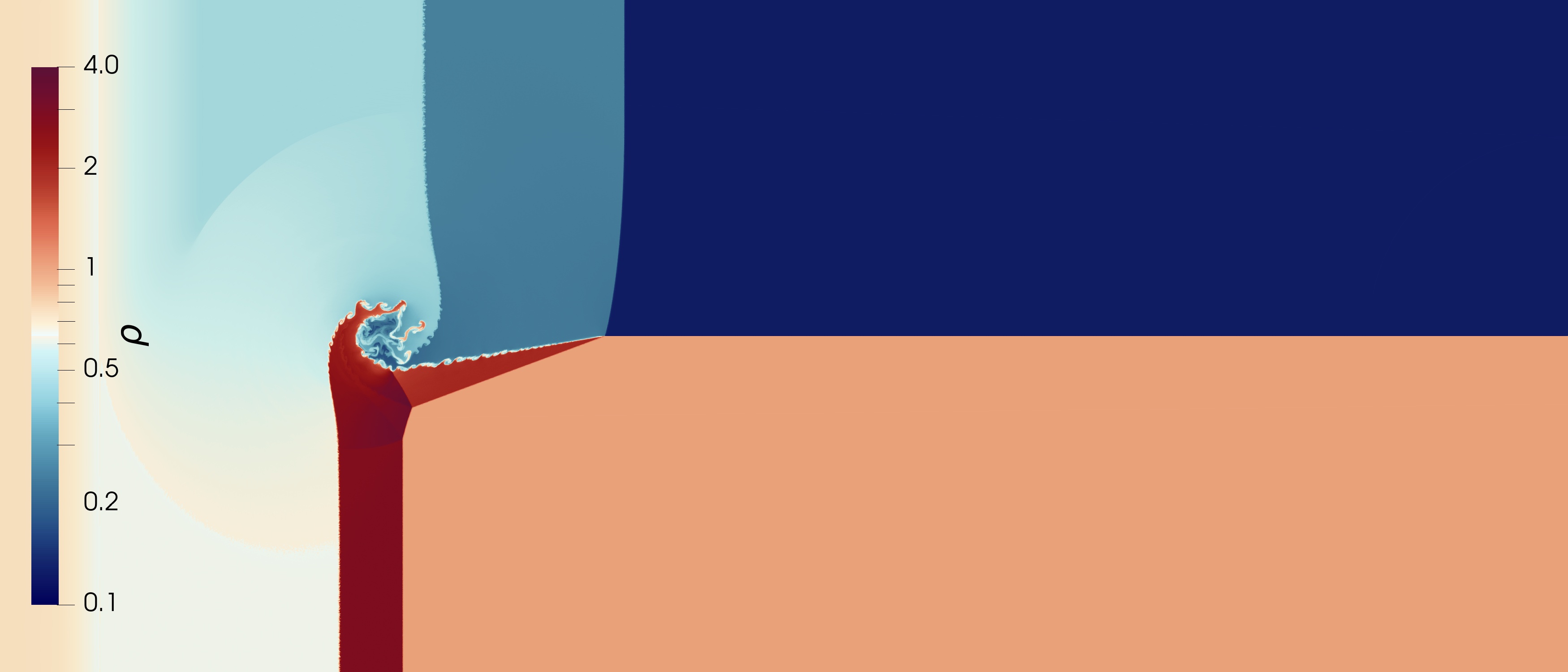}}}
        ~
        \subfloat[$t = 4$]{\adjustbox{width=0.47\linewidth, valign=b}{\includegraphics{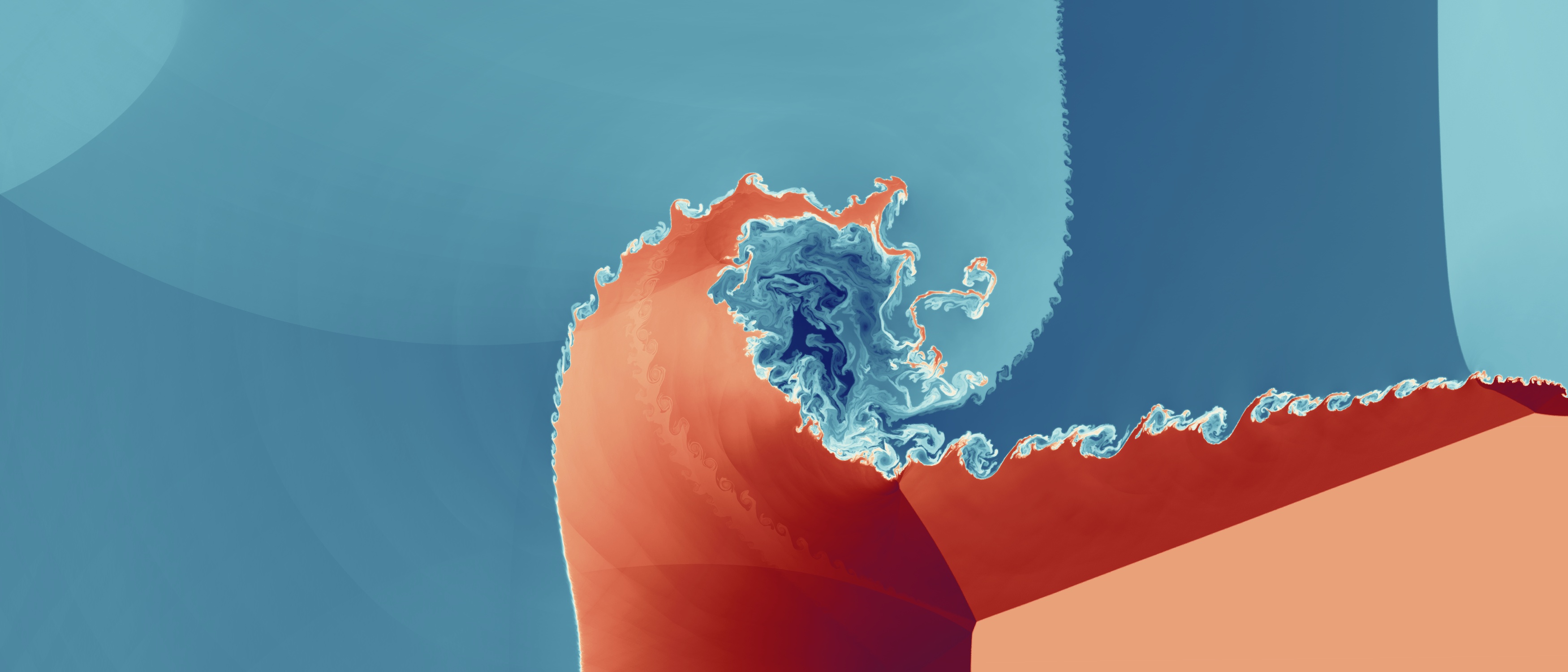}}}
        \newline
        \subfloat[$t = 2$]{\adjustbox{width=0.47\linewidth, valign=b}{\includegraphics{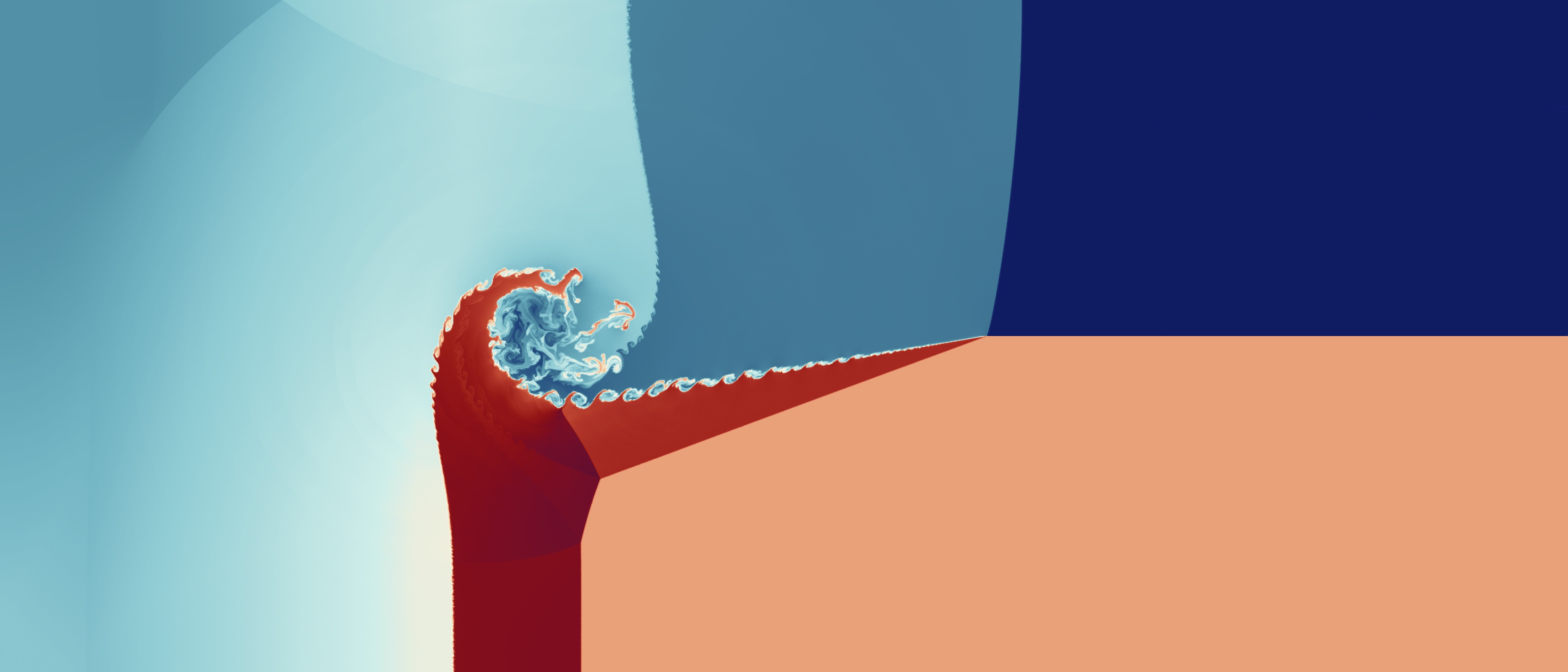}}}
        ~
        \subfloat[$t = 5$]{\adjustbox{width=0.47\linewidth, valign=b}{\includegraphics{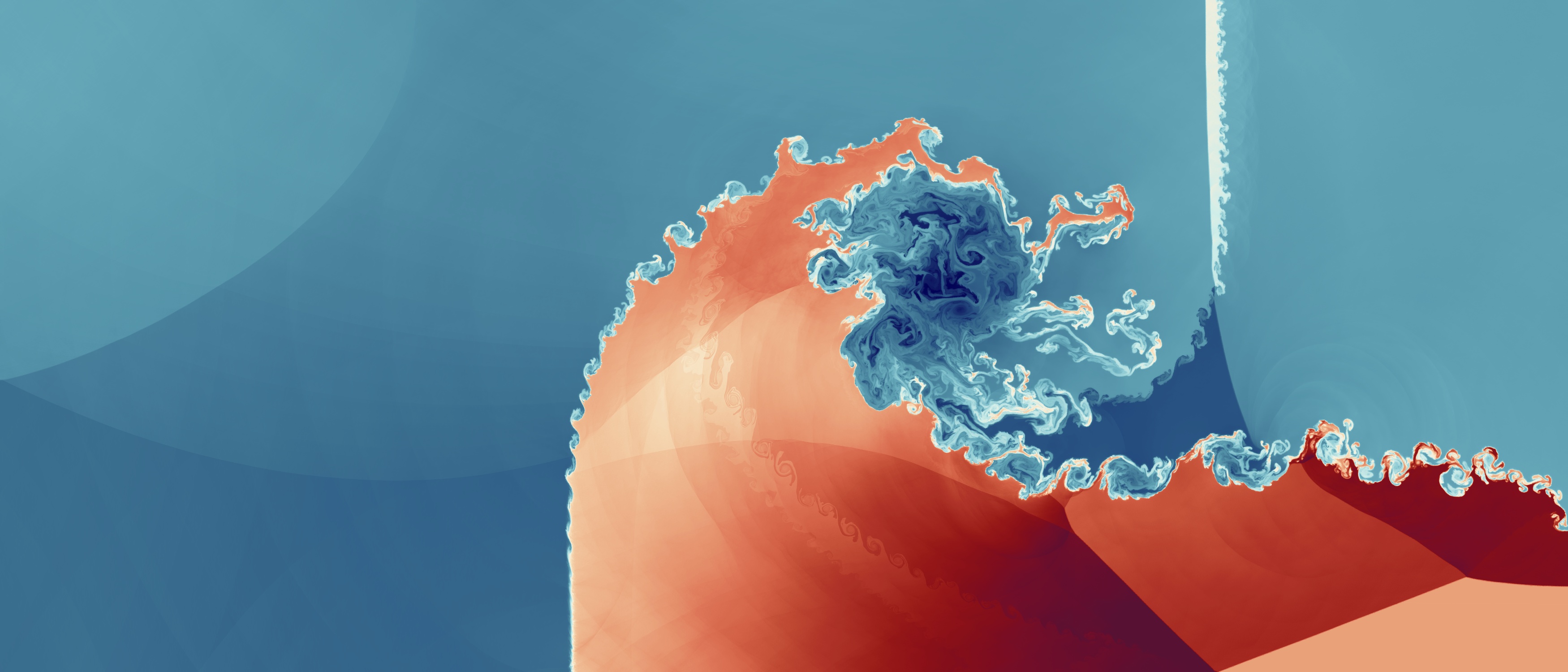}}}
        \newline
        \subfloat[$t = 3$]{\adjustbox{width=0.47\linewidth, valign=b}{\includegraphics{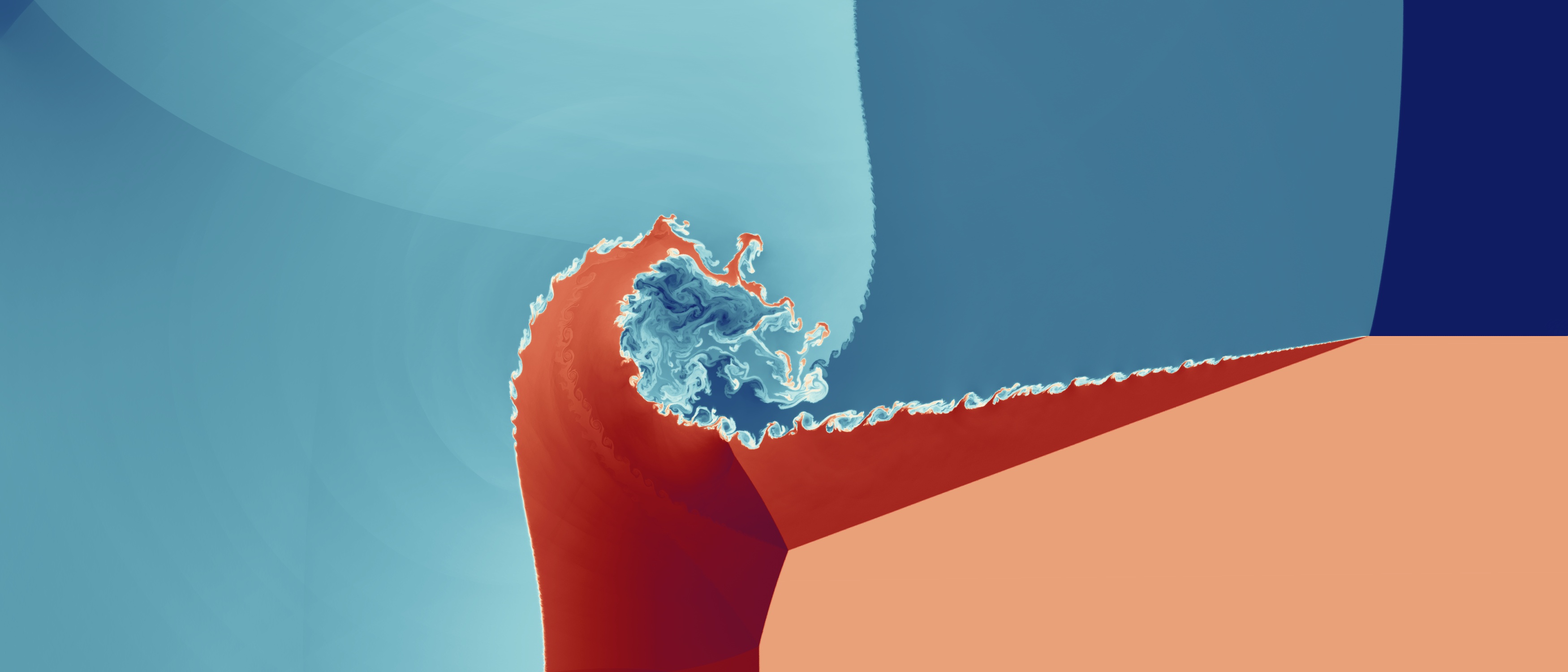}}}
        ~
        \subfloat[$t = 6$]{\adjustbox{width=0.47\linewidth, valign=b}{\includegraphics{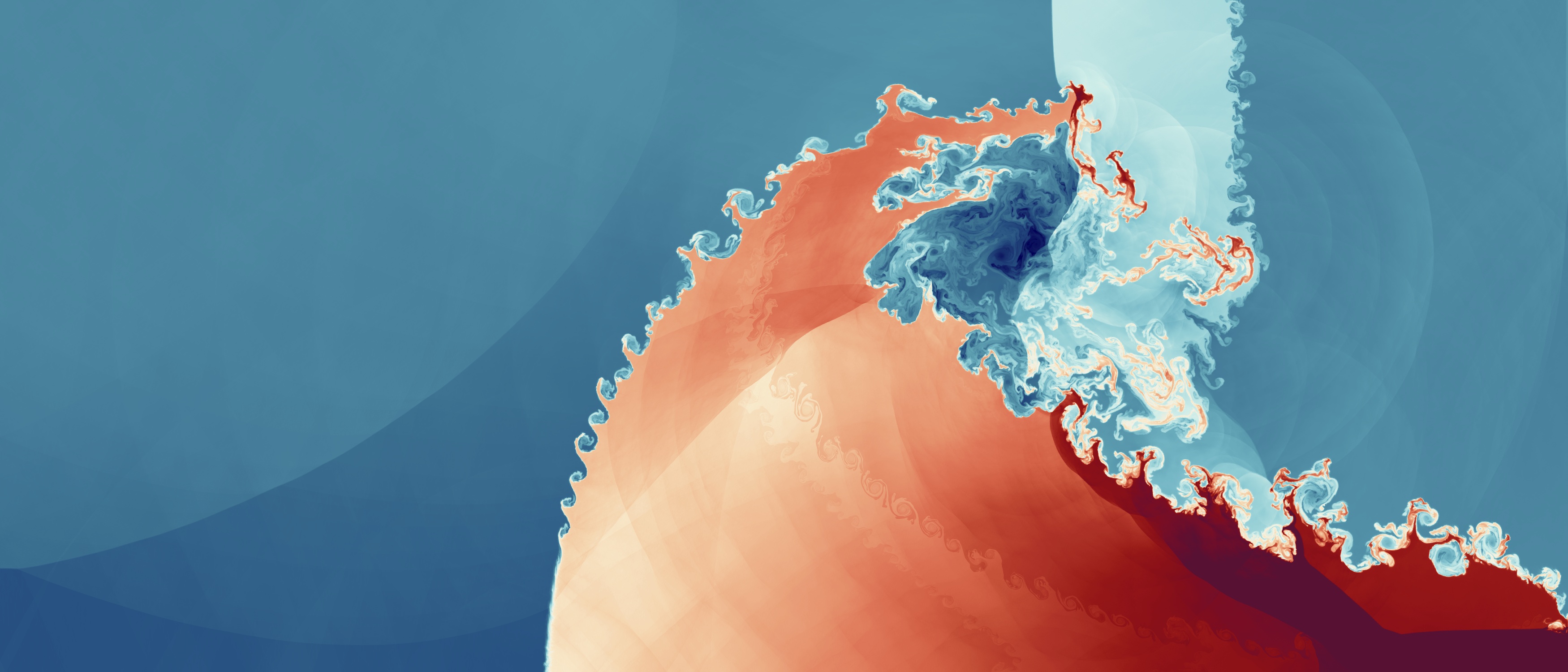}}}
        \newline
        \caption{\label{fig:triplepoint}Contours of density for the triple-point shock interaction problem at varying times computed using a $\mathbb P_3$ approximation on a $N = 2800 \times 1200$ structured quadrilateral mesh.}
    \end{figure}

The problem was solved using a $\mathbb P_3$ approximation on a uniform quadrilateral mesh with $N = 2800 \times 1200$ elements. The contours of mixture density at evenly-spaced time intervals are shown in \cref{fig:triplepoint}. It can be seen that the rollup of the contact discontinuity introduces Kelvin--Helmholtz instabilities in the flow which were well-resolved by the proposed scheme. The interaction of these vortical structures with the reflecting shock waves in the domain showcases the ability of the scheme to stabilise the solution around discontinuities without excessively dissipating small-scale flow structures. These results compare favourably to works such as that of \citet{Kolev2009} in terms of the resolving capability of small-scale flow structures.

\subsection{Multi-species Navier--Stokes equations}

\subsubsection{Rayleigh--Taylor instability}
    The extension to the multi-species viscous flow equations was performed through simulations of the Rayleigh--Taylor instability, a fundamental phenomenon stemming from the interaction of fluids of different densities in hydrostatic equilibrium. The problem consists of a denser fluid resting above a lighter fluid, with gravitational forces initially balanced by an equivalent pressure gradient. The interface between the fluids forms instabilities manifesting as ``fingers'' of heavy fluid descending in conjunction with ``pillars'' of lighter fluid ascending, driving the flow into a chaotic mixing state. 

    The problem is solved on the domain $\Omega = [-0.5, 0.5]^2 \times [-1, 1]$. The fluid interface was placed at $z = 0$, with $\alpha_0 \rho_0 = \rho_b$, $\alpha_1 \rho_1 = 0$ for $z \leq 0$ and $\alpha_0 \rho_0 = 0$, $\alpha_1 \rho_1 = \rho_t$ for $z > 0$. The densities of the light and heavy fluid were taken as $\rho_b = 1$ and $\rho_t = 3$, respectively, yielding an Atwood number of $1/2$. The fluid was taken to be at rest in the transverse directions, i.e., $u = v = 0$, and a small vertical velocity perturbation was imposed to seed instabilities in the flow. A deterministic single-mode sinusoidal perturbation was used with a cosinusoidal vertical decay as
    \begin{equation}
        w = A\cos \left (\frac{\pi z}{2L} \right)\sin \left (\frac{4\pi x}{L} \right)\sin \left (\frac{4\pi y}{L} \right),
    \end{equation}
    where the amplitude of the perturbation was taken as $A = 0.05$. Constant source terms of $-\rho g$ and $-\rho g w$, where $g = 1$, in the vertical momentum and energy equations, respectively, were used to simulate a gravitational field effect. The initial pressure field was chosen to enforce hydrostatic equilibrium in the flow with respect to the gravitational field, i.e.,
    \begin{equation}
        P = P_0 - \rho g z.
    \end{equation}
    The ambient pressure was set as $P_0 = 6$.

    A uniform hexahedral mesh with $N = 32 \times 32 \times 64$ elements was used with periodicity enforced along the transverse directions, and a slip adiabatic wall boundary condition was applied to the top and bottom walls. The specific heat constants for the lighter and heavier fluids were taken as $c_{v,0} = 3.11$, $c_{p,0} = 5.19$, $c_{v,1} = 0.72$, and $c_{p,1} = 1.007$. A uniform Prandtl number and dynamic viscosity were used, with the Prandtl number fixed at $Pr = 0.71$ and the dynamic viscosity fixed at $\mu = 1{\cdot}10^{-4}$, $5{\cdot}10^{-5}$, and $1{\cdot}10^{-5}$ for the varying simulations. 
    
    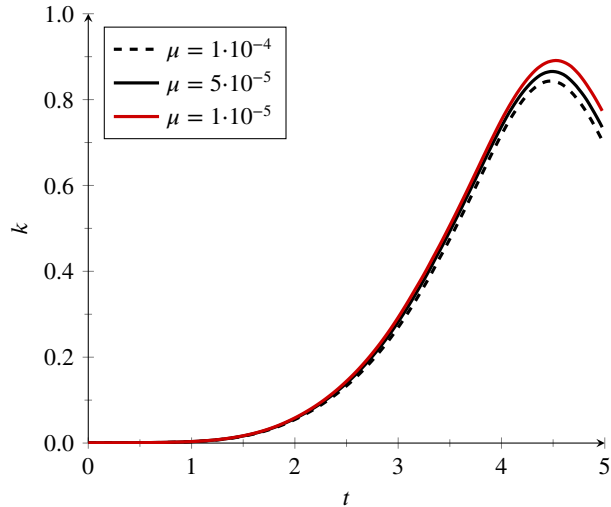
\begin{figure}[tbhp]
        \centering{\adjustbox{width=0.5\linewidth, valign=b}{\begin{tikzpicture}[spy using outlines={rectangle, height=3cm,width=2.3cm, magnification=3, connect spies}]
    \begin{axis}
    [
        axis line style={latex-latex},
        axis y line=left,
        axis x line=left,
        xmode=linear,
        ymode=linear,
        xlabel = {$t$},
        ylabel = {$k$},
        xmin = 0, xmax = 5,
        ymin = 0, ymax = 1,
        minor x tick num=1,
        minor y tick num=1,
        legend cell align={left},
        legend style={at={(0.03, 0.97)},anchor=north west},
        %axis line style={draw=none},
        %tick style={draw=none},
        x tick label style={/pgf/number format/.cd, fixed, fixed zerofill, precision=0, /tikz/.cd},
        y tick label style={/pgf/number format/.cd, fixed, fixed zerofill, precision=1, /tikz/.cd},
    ]
        
        \addplot[color=black, style={dashed, very thick}] table[x=t, y=k, col sep=comma]{./figs/data/rayleightaylor_mu1e-4.csv};
        \addlegendentry{$\mu = 1{\cdot}10^{-4}$}
        
        \addplot[color=black, style={very thick}] table[x=t, y=k, col sep=comma]{./figs/data/rayleightaylor_mu5e-5.csv};
        \addlegendentry{$\mu = 5{\cdot}10^{-5}$}
        
        \addplot[color=red!80!black, style={very thick}] table[x=t, y=k, col sep=comma]{./figs/data/rayleightaylor_mu1e-5.csv};
        \addlegendentry{$\mu = 1{\cdot}10^{-5}$}
        
    \end{axis}
\end{tikzpicture}}}
        \caption{\label{fig:rt_ke} Volume-averaged kinetic energy for the Rayleigh--Taylor instability problem computed using a $\mathbb P_3$ approximation on a $N = 32 \times 32 \times 64$ structured hexahedral mesh with varying values of the dynamic viscosity.}
    \end{figure}
    
    The problem was solved using a $\mathbb P_3$ approximation on the given mesh over the time range $t \in [0,5]$. The quantity of comparison for the flow was the volume-averaged kinetic energy, computed as
    \begin{equation}
        k = \frac{1}{2} \rho \mathbf{V}{\cdot}\mathbf{V}.
    \end{equation}
    A comparison of the kinetic energy profiles with respect to time for the varying values of the dynamic viscosity is shown in \cref{fig:rt_ke}. The instabilities driven by the gravitational field introduce kinetic energy into the initially static flow field. This forcing is eventually balanced by the viscous dissipation of the turbulent mixing state at $t \approx 4.5$, after which viscosity overcomes and starts driving down the kinetic energy of the system. It can be seen that with decreasing values of the dynamic viscosity, the kinetic energy peak in the flow increases. Given the identical mesh used for the simulations, this indicates that the dissipation in the flow is primarily introduced through the physical viscosity, not the numerical stabilisation approach, which indicates that the proposed method may be a promising approach for accurately simulating complex multi-species turbulent flows. 
    
    \begin{figure}[htbp!]
        \centering
        \subfloat[$t = 2$]{
        \adjustbox{width=0.24\linewidth,valign=b}{\includegraphics[width=\textwidth]{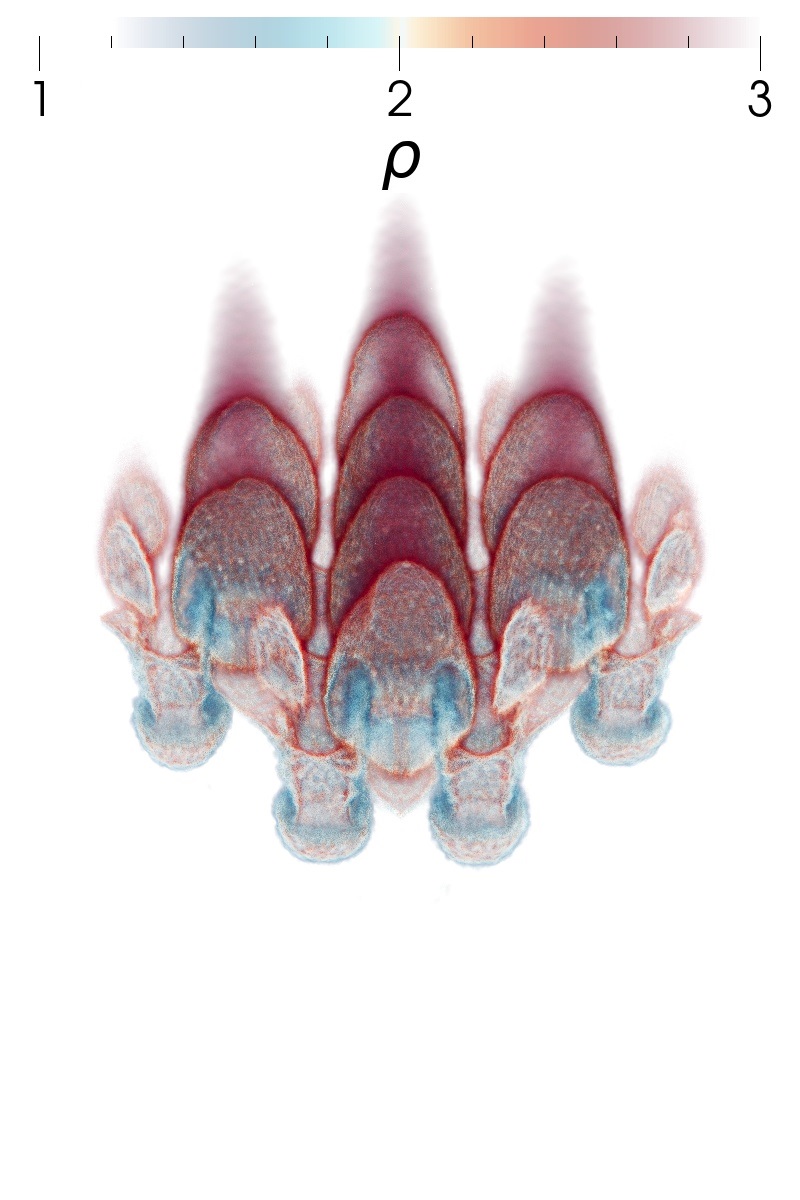}}}
        ~
        \subfloat[$t = 3$]{
        \adjustbox{width=0.24\linewidth,valign=b}{\includegraphics[width=\textwidth]{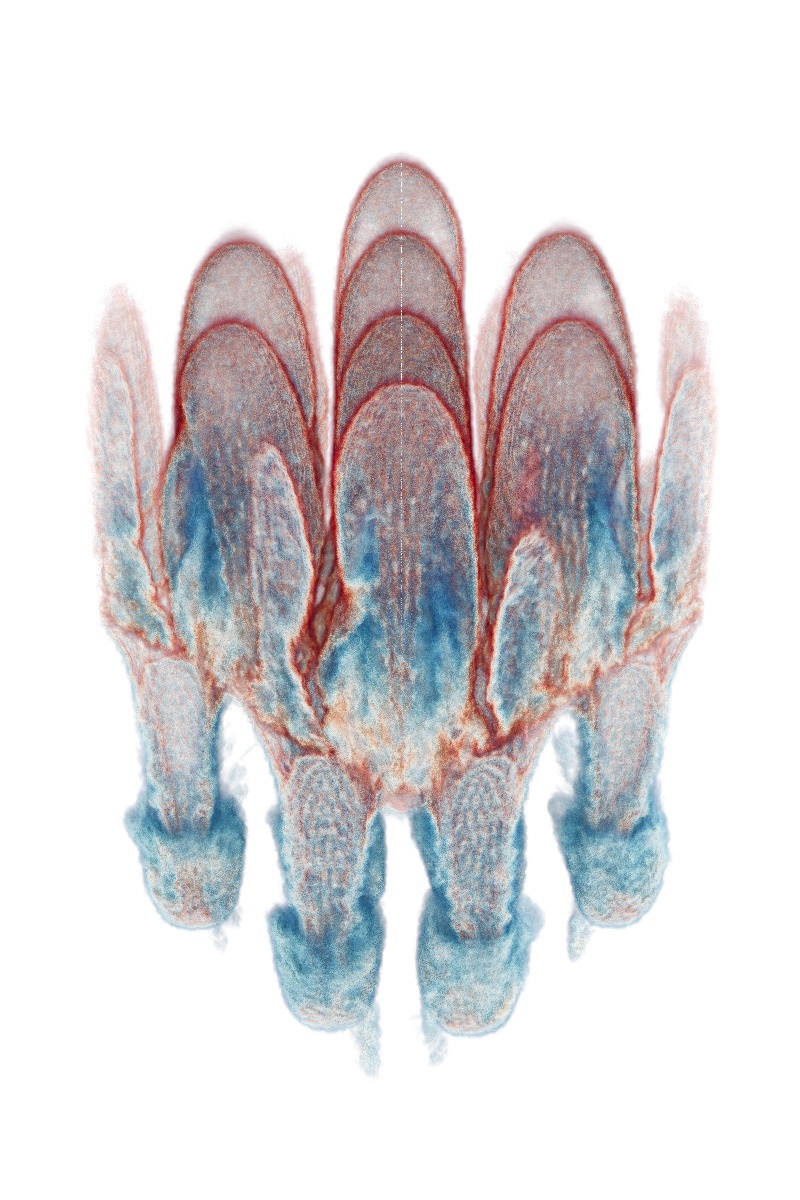}}}
        ~
        \subfloat[$t = 4$]{
        \adjustbox{width=0.24\linewidth,valign=b}{\includegraphics[width=\textwidth]{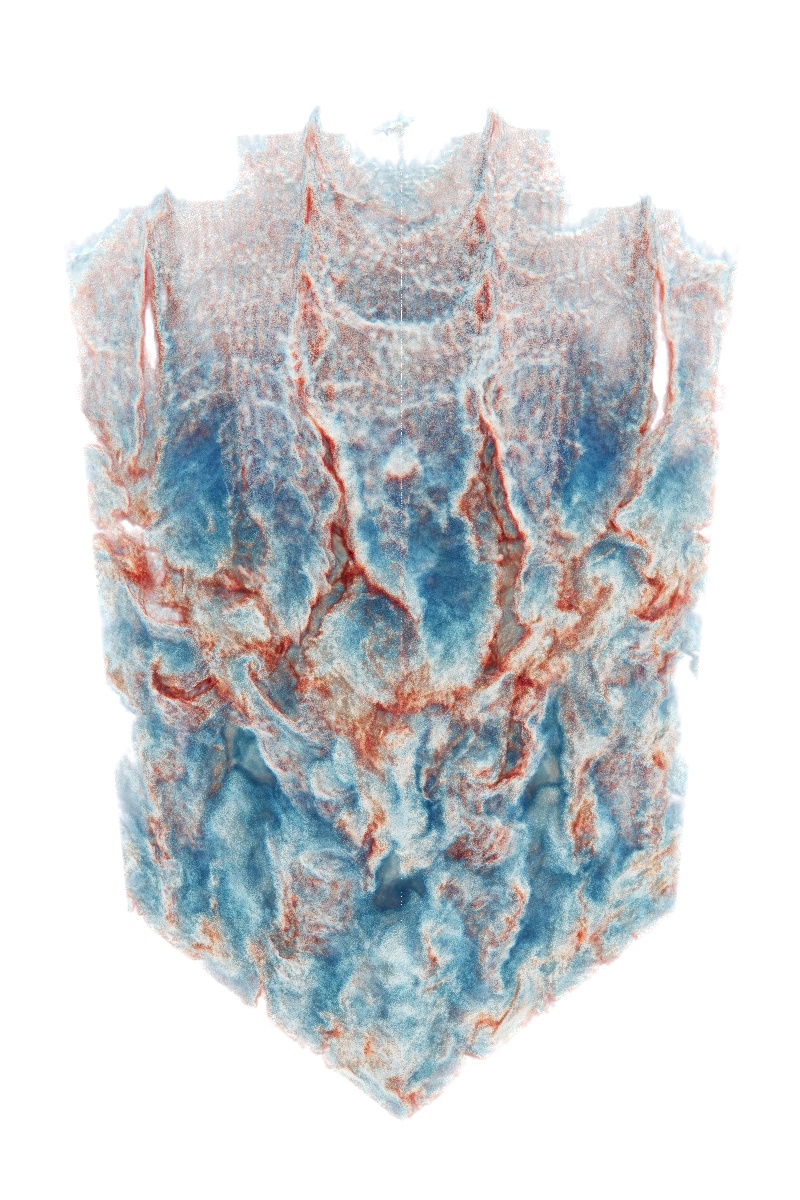}}}
        ~
        \subfloat[$t = 5$]{
        \adjustbox{width=0.24\linewidth,valign=b}{\includegraphics[width=\textwidth]{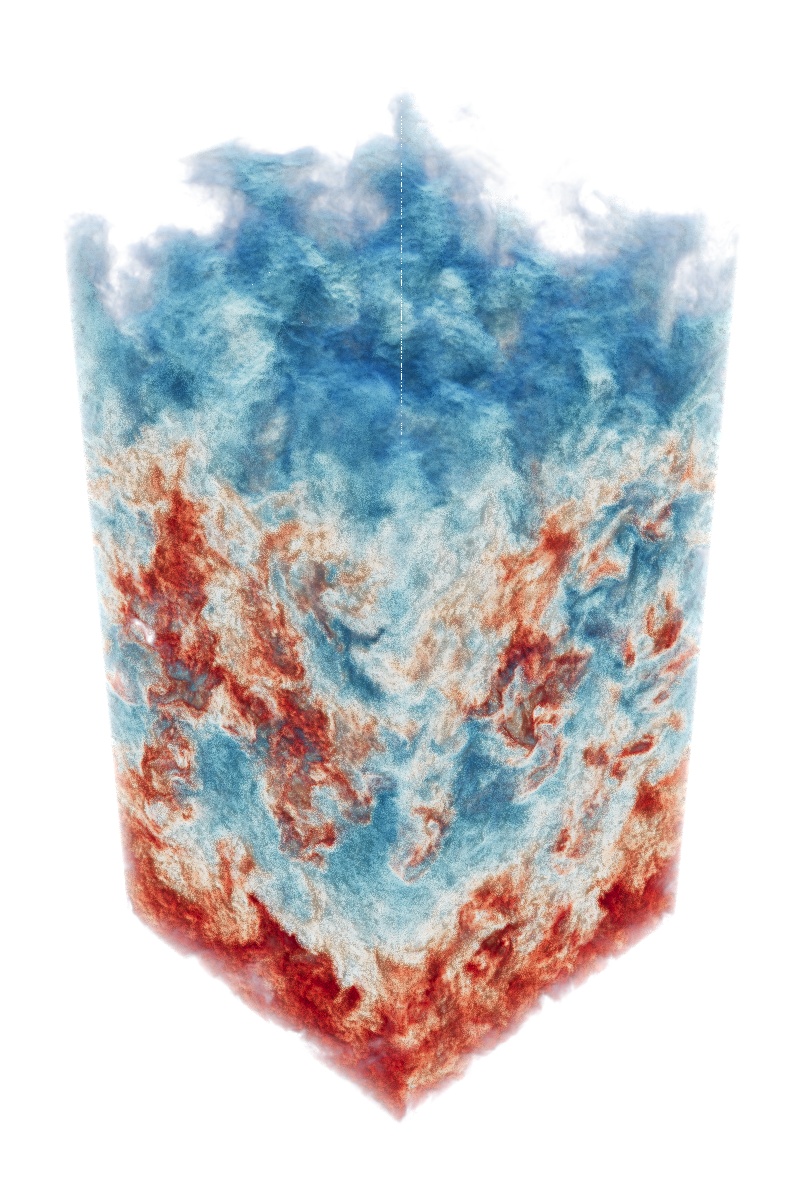}}}
        \newline
        \caption{\label{fig:rt_vis} Volume rendering of the density field for the Rayleigh--Taylor instability problem at varying times computed using a $\mathbb P_3$ approximation on a $N = 32 \times 32 \times 64$ structured hexahedral mesh with $\mu = 1{\cdot}10^{-5}$.}
    \end{figure}

    A visualisation of the flow through a volume rendering of the density field is also shown in \cref{fig:rt_vis} for the case of $\mu = 1{\cdot}10^{-5}.$ The canonical flow structures of the descending ``fingers'' and ascending ``pillars'' of the Rayleigh--Taylor instability can be clearly seen in the early stages of the flow. The breakdown of these flow structures can then be observed at $t = 4$, after which a chaotic turbulent mixing state develops at $t = 5$. Eventually, the viscous dissipation in the system is expected to gradually bring the flow to a rest over longer simulation times.
    
\subsubsection{Taylor--Green vortex}
    The final evaluation of the proposed method was performed on the Taylor--Green vortex, a canonical fluid dynamics problem for studying vortex dynamics and turbulent transition and decay. The problem consists of a laminar initial flow field which transitions to turbulence at Reynolds numbers of approximately 500 and above, which makes it an ideal case for evaluating the effects of the proposed scheme on predicting small-scale turbulent flow structures, particularly with respect to the numerical dissipation introduced by the scheme to stabilise the solution around species interfaces. To comprehensively study this, we consider the case of a two-species flow with identical specific heat constants and transport coefficients for both species which should, in theory, exhibit identical flow features to the single-species case. The resulting differences between the two cases can then be attributed to effects of the proposed numerical stabilisation method. 

    For this problem, the domain is taken as $\Omega = [0, 2\pi]^3$, and the initial conditions are defined as 
    \begin{subequations}
        \begin{align}
            \rho &= 1,\\
            u &= \sin{x}\cos{y}\cos{z}, \\
            v &= -\cos{x}\sin{y}\cos{z}, \\
            w &= 0,\\
            P &= P_0 + \frac{1}{16}\left(\cos{(2x)} + \cos{(2y)}\right)(\cos{(2z + 2)}),
        \end{align}
    \end{subequations}
    where the ambient pressure was set as $P_0=1/(\gamma M^2)$ for a specific heat ratio of $\gamma=1.4$ and a Mach number of $M = 0.08$. The specific heat capacities of both fluid were set as $c_p=1.005$ and $c_v=0.718$. The Reynolds number and Prandtl number were set as $Re = 1600$ and $Pr = 0.71$, respectively. At these operating conditions, the Taylor--Green vortex is an extensively studied case~\citep{Brachet1983}.
    
    \begin{table}[tbhp]
        \centering
        \caption{\label{tab:tgv_cases} Overview of the various problem setups for the Taylor--Green vortex.}
        \begin{tabular}{l r r r}
            \toprule
            Test & Species ($n$) & $\alpha_0$ & $N$ \\ \midrule
            A-coarse & 1 & 1 & $32^3$ \\
            A-fine & 1 & 1 & $40^3$ \\
            B-coarse & 2 & \cref{eq:a0_smooth} & $32^3$ \\
            B-fine & 2 & \cref{eq:a0_smooth} & $40^3$ \\
            C-coarse & 2 & \cref{eq:a0_sharp} & $32^3$ \\
            C-fine & 2 & \cref{eq:a0_sharp} & $40^3$ \\
            \bottomrule
        \end{tabular}
    \end{table}
    
    Three variations of the problem were considered. The first variation, denoted by the label $A$, consists of the standard single-species description of the problem, where $\rho = \rho_0$. The second variation, denoted by the label $B$, consists of a two-species flow with an initially smooth species interface, given as
    \begin{equation}\label{eq:a0_smooth}
        \alpha_0 = \half  + \frac{1}{16}(\cos{(x)} + 1)(\cos{(y)} + 1)(\cos{(z)} + 1).
    \end{equation}
    The third variation, denoted by the label $C$, consists of a two-species flow with an initially sharp species interface, given as    
    \begin{equation}\label{eq:a0_sharp}
        \alpha_0= \half + \frac{1}{4}\tanh{\big(\beta(x - \pi)\big)}\tanh{\big(\beta(y - \pi)\big)}\tanh{\big(\beta(z - \pi)\big)},
    \end{equation}
    where $\beta = 1000$. For each variation of the problem, both a coarse mesh, consisting of $N = 32^3$ elements, and a fine mesh, consisting of $N = 40^3$ elements, were considered. An overview of these problem setups is presented in \cref{tab:tgv_cases}. 

    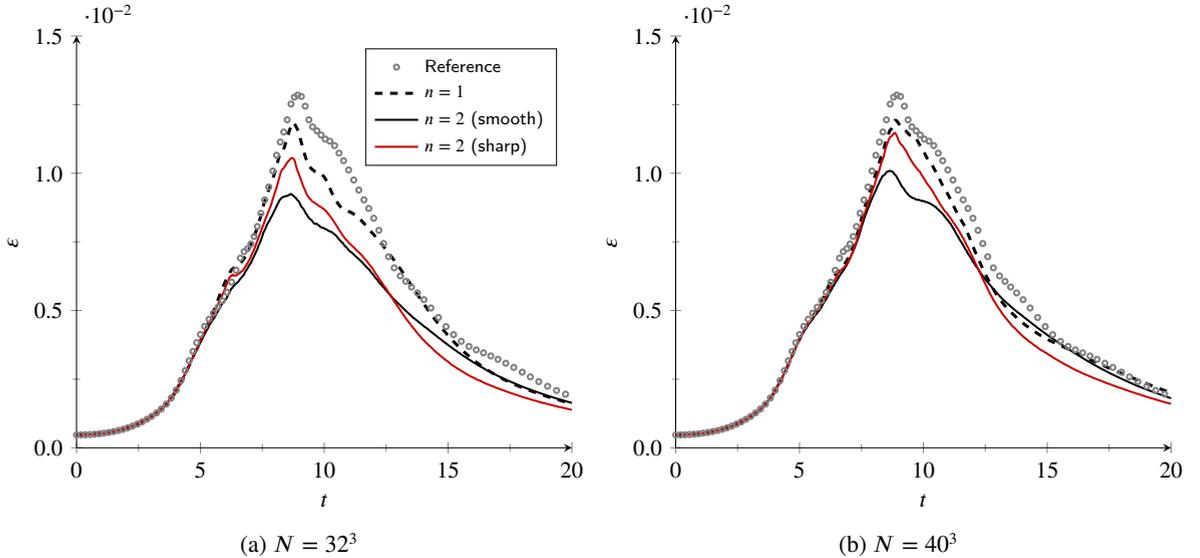
\begin{figure}[tbhp]
        \centering
        \subfloat[$N = 32^3$]{{\adjustbox{width=0.48\linewidth, valign=b}{\begin{tikzpicture}[spy using outlines={rectangle, height=3cm,width=2.3cm, magnification=3, connect spies}]
    \begin{axis}
    [
        axis line style={latex-latex},
        axis y line=left,
        axis x line=left,
        xmode=linear,
        ymode=linear,
        xlabel = {$t$},
        ylabel = {$\varepsilon$},
        xmin = 0, xmax = 20,
        ymin = 0, ymax = 0.015,
        minor x tick num=1,
        minor y tick num=1,
        legend cell align={left},
        legend style={font=\scriptsize, at={(0.97, 0.97)},anchor=north east},
        %axis line style={draw=none},
        %tick style={draw=none},
        x tick label style={/pgf/number format/.cd, fixed, fixed zerofill, precision=0, /tikz/.cd},
        y tick label style={/pgf/number format/.cd, fixed, fixed zerofill, precision=1, /tikz/.cd},
    ]
        \addplot[ color=gray, style={thick}, only marks, mark=o, mark options={scale=0.5}, mark repeat = 20, mark phase = 0] table[x=t, y=dkep, col sep=comma, mark=*]{./figs/data/van_rees_2011_dkep_PSP512.csv};
        \addlegendentry{Reference}
    
        \addplot[color=black, style={dashed, very thick}] table[x=t, y expr=\thisrow{sp_32}, col sep=comma]{./figs/data/tgv_merged_rr_et1e-4_data.csv};
        \addlegendentry{$n = 1$}
        
        \addplot[color=black, style={thick}] table[x=t, y expr=\thisrow{mp_sm_32}, col sep=comma]{./figs/data/tgv_merged_rr_et1e-4_data.csv};
        \addlegendentry{$n = 2$ (smooth)}
        
        \addplot[color=red!80!black, style={thick}] table[x=t, y expr=\thisrow{mp_sh_32}, col sep=comma]{./figs/data/tgv_merged_rr_et1e-4_data.csv};
        \addlegendentry{$n = 2$ (sharp)}
        
    \end{axis}
\end{tikzpicture}}}}
        \subfloat[$N = 40^3$]{{\adjustbox{width=0.48\linewidth, valign=b}{\begin{tikzpicture}[spy using outlines={rectangle, height=3cm,width=2.3cm, magnification=3, connect spies}]
    \begin{axis}
    [
        axis line style={latex-latex},
        axis y line=left,
        axis x line=left,
        xmode=linear,
        ymode=linear,
        xlabel = {$t$},
        ylabel = {$\varepsilon$},
        xmin = 0, xmax = 20,
        ymin = 0, ymax = 0.015,
        minor x tick num=1,
        minor y tick num=1,
        legend cell align={left},
        legend style={font=\scriptsize, at={(0.97, 0.97)},anchor=north east},
        %axis line style={draw=none},
        %tick style={draw=none},
        x tick label style={/pgf/number format/.cd, fixed, fixed zerofill, precision=0, /tikz/.cd},
        y tick label style={/pgf/number format/.cd, fixed, fixed zerofill, precision=1, /tikz/.cd},
    ]
        \addplot[ color=gray, style={thick}, only marks, mark=o, mark options={scale=0.5}, mark repeat = 20, mark phase = 0] table[x=t, y=dkep, col sep=comma, mark=*]{./figs/data/van_rees_2011_dkep_PSP512.csv};
        % \addlegendentry{Reference}
        
        \addplot[color=black, style={dashed, very thick}] table[x=t, y expr=\thisrow{sp_40}, col sep=comma]{./figs/data/tgv_merged_rr_et1e-4_data.csv};
        % \addlegendentry{Single species}
        
        \addplot[color=black, style={thick}] table[x=t, y expr=\thisrow{mp_sm_40}, col sep=comma]{./figs/data/tgv_merged_rr_et1e-4_data.csv};
        % \addlegendentry{Multi-species (smooth)}

        \addplot[color=red!80!black, style={thick}] table[x=t, y expr=\thisrow{mp_sh_40}, col sep=comma]{./figs/data/tgv_merged_rr_et1e-4_data.csv};
        % \addlegendentry{Multi-species (sharp)}
        
    \end{axis}
\end{tikzpicture}}}}

        \caption{\label{fig:tgv} Dissipation measured by enstrophy for the Taylor--Green vortex computed using a $\mathbb P_3$ approximation on a $N = 32^3$ (left) and $N = 40^3$ (right) structured hexahedral mesh. Results shown for problem setups with single-species (black, dashed), multi-species with smooth phase boundaries (black, solid), and multi-species with sharp phase boundaries (red, solid). DNS results of \citet{vanRees2011} (private communication) shown for reference.}
    \end{figure}
    
    The metric of interest in the flow was the dissipation rate measured by the enstrophy, defined as
    \begin{equation}
        \varepsilon = \frac{2\mu}{(2\pi)^3}\int_{\Omega}\half\rho(\pmb{\omega}\cdot\pmb{\omega})\mathrm{d}\mathbf{x}, 
    \end{equation}
    where $\pmb{\omega}=\boldsymbol{\nabla}\times\mathbf{V}$ is the vorticity. As this functional is based on the vorticity in the flow field, an underprediction in the enstrophy indicates that the generation of small-scale flow features has been suppressed by numerical dissipation. The problem was solved using a $\mathbb P_3$ approximation, and the dissipation measured by enstrophy for the various problem setups on the coarse and fine mesh is presented in \cref{fig:tgv} in comparison to the direct numerical simulation (DNS) results of \citet{vanRees2011} (obtained via private communication). It can be seen that the single-species results ($A$) show relatively good agreement with the DNS results in terms of enstrophy for the coarse mesh and show convergence with increasing resolution. When the flow field was described using a two-species approach ($B$ and $C$), the measured enstrophy was slightly underpredicted with respect to the single-species approach. However, it was observed that these differences diminished with increasing resolution, with the case of the initially sharp interface ($C$) showing relatively good agreement with the single-species case. Notably, the case of the initially smooth interface ($B$) underpredicted the enstrophy in the flow more than the initially sharp interface, indicating that more numerical dissipation was introduced for a smoothly varying interface than a sharp interface. While this may seem counter-intuitive at first as it is expected that the numerical dissipation introduced by the proposed approach is primarily for the purpose of stabilising the scheme in the vicinity of discontinuities, these observations may be linked to aliasing errors associated with the nonlinear description of the interface. It has been shown that the enforcement of entropy constraints tends to mitigate aliasing errors \citep{Dzanic2023} through the addition of a suitable amount of numerical dissipation, which may explain the observed underprediction of the enstrophy. Regardless, the results indicate that the proposed approach can accurately resolve complex flow phenomena such as transition to turbulence for multi-species flows to a similar degree of accuracy that can be obtained by high-order DSEM for single-species flows. 
    
\section{Conclusions}\label{sec:conclusions}
    In this work, we present a novel positivity-preserving numerical stabilisation approach for the conservative compressible multi-species flow equations solved via high-order discontinuous spectral element methods. The stabilisation method uses the adaptive nonlinear filtering approach of \citet{Dzanic2022}, with constraints applied to species density, total density, and pressure. It was found that the entropy constraints proposed in \citet{Dzanic2022} led to excessive dissipation around species interfaces, substantially degrading the accuracy of the underlying high-order scheme. An improved stabilisation framework was proposed in this work where the entropy component of the constraints was adaptively enabled or disabled based on the presence of discontinuities in the pressure field, detected using a parameter-free sensor which utilises the convergence properties of high-order DSEM. The efficacy of the proposed method was demonstrated in numerical experiments on the multi-species Euler and Navier--Stokes equations computed on structured and unstructured grids. It was shown that the scheme can recover the high-order accuracy of the underlying DSEM for smooth solutions while ensuring robustness in the vicinity of shock waves and contact discontinuities, with promising results in the ability to predict nonlinear flow phenomena such as flow instabilities and shock-vortex interactions. Furthermore, the proposed modification of the entropy constraints showed significant improvements in the ability of the method to accurately resolve species interfaces. Utilising the ability of the proposed framework to robustly and accurately resolve multi-species fluid flows, future work will focus on extending the approach to nonlinear equations of state and reacting multi-species flows.

\printcredits
\section*{Acknowledgements}
WT would like to thank Robert Manson-Sawko for his insightful discussions. WT also acknowledges the support of UKRI through the grant MR/T041862/1 and the IBM Research Cognitive Computing Cluster service for providing resources that have contributed to the research results reported within this paper. TD would like to acknowledge the computational resources provided by the Princeton Institute for Computational Science and Engineering. 

\bibliographystyle{cas-model2-names}
\bibliography{references}

\end{document}